\pdfoutput=1
\documentclass[
    floatfix,
    preprint,
    showpacs,
    preprintnumbers,
    amsmath,
    amssymb
]{revtex4-2}
\usepackage[utf8]{inputenc}

\usepackage[USenglish]{babel}

\usepackage[dvipsnames]{xcolor}

\usepackage{accents}
\usepackage{amsmath}
\usepackage{amssymb}
\usepackage{bbold}
\usepackage{bm}
\usepackage[inline]{enumitem}
\usepackage{graphicx}
\usepackage{hyperref}
\hypersetup{pdfencoding=auto,psdextra}
\usepackage{textgreek}
\usepackage{bookmark}
\usepackage{cleveref}
\usepackage{fancyvrb}
\usepackage{siunitx}
\usepackage{tabularx}
\usepackage{multirow}
\usepackage{xspace}
\usepackage{url}

\newcommand{\ubar}[1]{\underaccent{\bar}{#1}}

\newcommand{\refe}[1]{Eq.~\ref{#1}}
\newcommand{\reff}[1]{Fig.~\ref{#1}}

\newcommand{\reft}[1]{Table~\ref{#1}}

\begin{document}

\title{Inverse modeling of circular lattices via orbit response measurements in the presence of degeneracy}

\author{D. Vilsmeier}\email{d.vilsmeier@gsi.de}
\affiliation{Johann Wolfgang Goethe-University Frankfurt, 60323 Frankfurt am Main, Germany}
\author{R. Singh}
\affiliation{GSI Helmholtz Centre for Heavy Ion Research, 64291 Darmstadt, Planckstr. 1, Germany}
\author{M. Bai}
\affiliation{SLAC National Accelerator Laboratory, 2575 Sand Hill Rd, Menlo Park, CA 94025, USA}

\date{\today}

\begin{abstract}
The number and relative placement of BPMs and steerers with respect to the quadrupoles in a circular lattice can lead to degeneracy in the context of inverse modeling of accelerator optics. Further, the measurement uncertainties introduced by beam position monitors can propagate by the inverse modeling process in ways that prohibit the successful estimation of model errors.
In this contribution, the influence of BPM and steerer placement on the conditioning of the inverse problem is studied. An analytical version of the Jacobian, linking the quadrupole gradient errors along with BPM and steerer gain errors with the orbit response matrix, is derived. It is demonstrated that this analytical version of the Jacobian can be used in place of the numerically obtained Jacobian during the fitting procedure. The approach is first tested with simulations and the findings are verified by measurement data taken on SIS18 synchrotron at GSI.
The results are crosschecked with the standard numerical Jacobian approach. The quadrupole errors causing tune discrepancies observed at SIS18 are identified.
\end{abstract}

\maketitle

\newpage
\tableofcontents
\newpage

\section{Introduction}

Precise knowledge of the lattice's optics elements is crucial for optimal operation  of any circular accelerator. It is especially important for the flexible and fast ramping synchrotron like SIS18 where transient effects can change the lattice properties cycle to cycle as well as during the energy ramp. Inability to identify these changes or model errors in general can lead to beam emittance dilution or beam losses. Linear Optics from Closed Orbits (LOCO) is a common method for machine model estimation which relies only on the measurement of the orbit response matrix (ORM). First detailed discussion of LOCO can be found in \cite{LOCO:1997} and since then the technique has experienced frequent usage at different institutes~\cite{LOCO:APP:SOLEIL,LOCO:APP:AustralianSynchrotron2007,LOCO:APP:AustralienSynchrotron2011,LOCO:APP:SwissLightSource}. Typically, LOCO takes a measured ORM and varies all relevant lattice parameters in a multi-dimensional optimization problem to match the simulated with the measured ORM. Based on the outcome of the optimization procedure, machine parameters are adjusted to reach the design values.

Beam position monitor (BPM) errors are unavoidable during measurement and will cast an uncertainty on the measured ORM. This uncertainty then propagates through the inverse modeling process and influences the precision of derived parameters. Depending on the lattice and the optics, the effect of BPM errors can be more or less problematic for the accuracy of inverse modeling results. In some cases, the influence of BPM errors can even hinder the successful reconstruction of quadrupole errors.
An improvement of the efficiency was introduced in~\cite{LOCO:2005:HuangThesis} by adding specific constraints for the fitting parameters. A related approach for improving the efficiency was introduced in~\cite{LOCO:2009:ImprovedFitting}.

Because measurement of the ORM typically varies one steerer at a time it can take significant amount of machine time. There have been efforts to reduce the time and impact of the measurement, for example by sine-wave excitation of multiple steerers at different frequencies simultaneously~\cite{LOCO:APP:MultifrequencyMode}. Another approach used the data obtained from closed-orbit feedback correction to continuously update an estimate of the ORM; for sufficient number of iterations, this will converge to the true ORM~\cite{LOCO:APP:ContinuousMeasurement}.
In addition to the measurement time, the inverse modeling process itself contributes to the required time until results are available. While different optimizers need different number of iterations until convergence, Jacobian-based optimizers use by far the fewest number of iterations since the Jacobian contains lots of information about where the minima lies. However, significant time is spent to compute the Jacobian via finite-difference approximation. One aspect of the presented work is to reduce the Jacobian's computation time.

In this contribution, we derive an analytical version of the Jacobian relating the ORM and the quadrupole strength errors along with BPM and steerer gain errors. This Jacobian matrix is used by the optimizer, e.g. Levenberg-Marquardt, in order to improve the current best guess of lattice errors during an iterative process. We have studied the properties of this analytical Jacobian with respect to conditioning of the inverse problem. We show that the analytical Jacobian highlights all relevant properties of the model error estimation problem. Rank deficiency of the Jacobian implies a degeneracy of the inverse problem while small eigenvalues of the Jacobian suggest quasi-degeneracy for some patterns of quadrupole errors. These patterns are more susceptible to measurement uncertainty. We further use the analytical version of the Jacobian, obtained from the lattice's Twiss data, during the fitting procedure and show that it reaches convergence similar to using the numerically obtained Jacobian. The analytical Jacobian is obtained quickly since it requires only a single Twiss computation for the lattice.

In light of the evaluated Jacobian and its discussed properties, inverse modeling of the SIS18 synchrotron is performed for the first time. We have identified and diagnosed several model errors for SIS18. A notable error is the tune offset of 0.02 units in the horizontal plane which was a known discrepancy for several years in the SIS18 machine model.

In this process, we also devised a general iterative method for automatic and online correction of quadrupolar errors simply based on the analytical Jacobian and measured ORM. This method has similarities with iterative closed orbit correction.

In the following, the structure of the paper is described.
In section~\ref{sec:orbit-response-matrix} we introduce the lattice used throughout this contribution and the concept of orbit response matrix.
Section~\ref{sec:degeneracy} explains the inverse problem with regard to degeneracy of its solutions. The analytical derivation of the Jacobian is presented. Also, the influence of BPM and steerer placement on the degeneracy is shown.
Section~\ref{sec:fitting-with-jacobian} discusses the fitting procedure by using the Jacobian as well as discusses the convergence properties for different approaches.
In section~\ref{sec:experimental-data} the experimental results are presented.

\section{Orbit response matrix}
\label{sec:orbit-response-matrix}

The orbit change $x_{b}$ at BPM $b$ when changing the steerers indexed with $s$ by a kick $\delta_s$, is given by~\cite{SYLee}:

\begin{equation}\label{eq:orbit-response}
    x_b = \sum_s\delta_s\left[\frac{\sqrt{\beta_b\beta_s}}{2\sin(\pi Q)}\cos(\pi Q - |\mu_b - \mu_s|) - \frac{D_b D_s}{\left(\frac{1}{\gamma^2} - \frac{1}{\gamma_t^2}\right)C}\right]
\end{equation}

where $\beta_{b,s}$ and $\mu_{b,s}$ denote, respectively, the beta functions and the phase advances at BPM and steerer position, and $Q$ is the betatron tune. In the second term, $D_{b,s}$ denotes the the dispersion at BPM and steerer position and $C$ is the circumference of the synchrotron; $\gamma$ and $\gamma_t$ denote, respectively, the beam energy and transition energy of the lattice ($\gamma = \frac{E}{E_0}$). This term is only relevant for synchrotrons operating near transition energy.

Hence, the orbit change is a linear function in the applied kick and it encodes the optics via the lattice functions $\beta$ and $\mu$. The orbit response $r_{bs}$ at BPM $b$ reacting to a single steerer $s$ is defined as:
\begin{equation}
    r_{bs} = \frac{x_b}{\delta_s}
\end{equation}

The orbit response matrix (ORM) arranges the orbit responses for all BPM/steerer pairs in a matrix form: $r_{bs}$ where $b$ is the row index and refers to BPMs and $s$ is the column index and refers to steerers.

\begin{figure}[hbt]
   \centering
   \includegraphics[width=.49\textwidth]{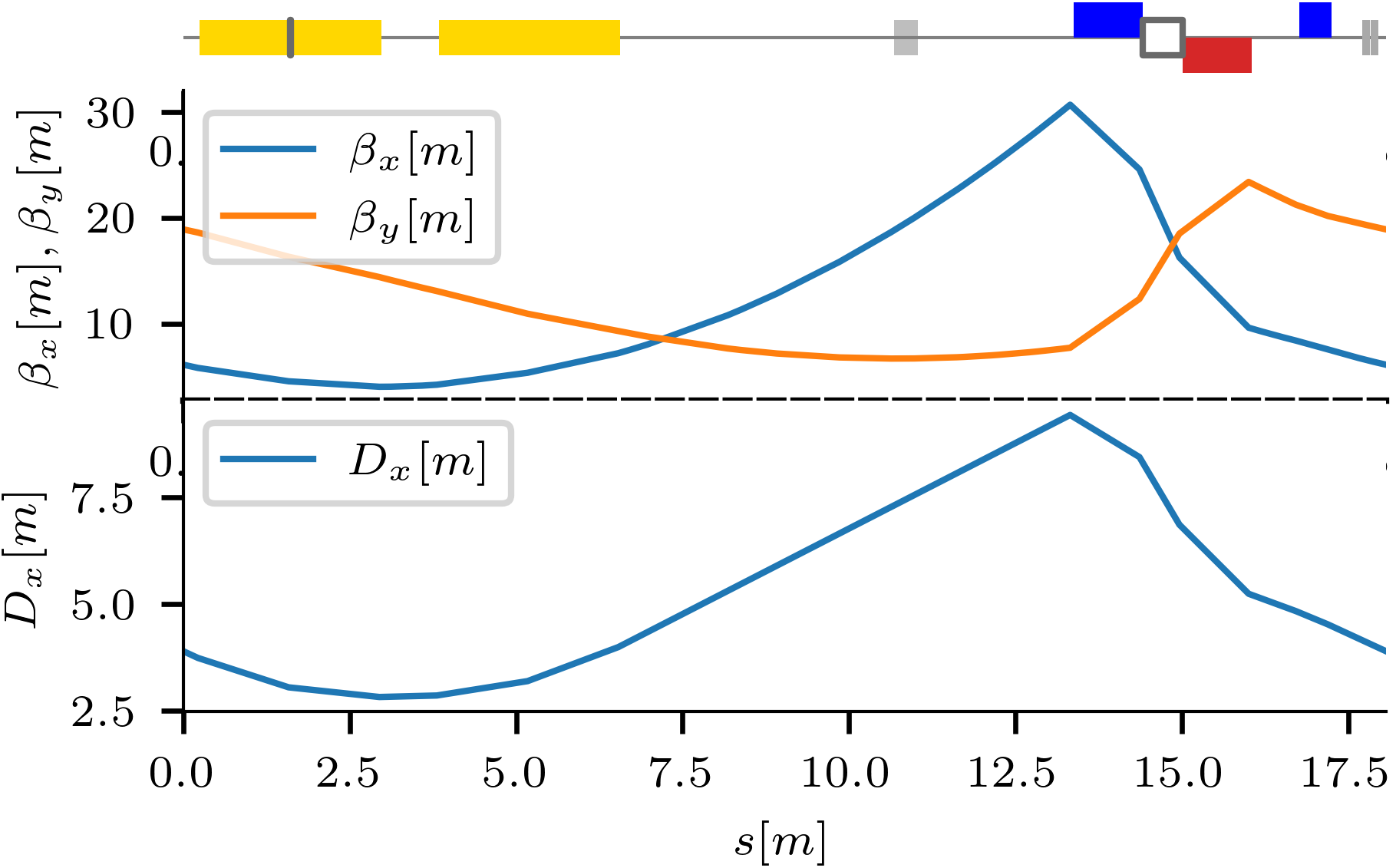}
   \caption{Schematic of SIS18 lattice with optics functions showing the first of in total twelve sections. The twelve sections are identical except that in sections 4 and 6 the horizontal steerer is located on the second bending magnet rather than the first. Blue (raised): focusing quadrupoles, red (lowered): defocusing quadrupoles, yellow (centered): bending magnets; the horizontal steerer is shown as a black line on top of the first bending magnet (in sections 4 and 6 it is located on the second bending magnet); the vertical steerer is shown as a gray box between the focusing and defocusing quadrupole; the vertical and horizontal BPMs (in that order) are shown as gray solid boxes downstream of the third quadrupole (since they are right next to each other, they might appear as a single gray box).}
   \label{fig:orm:sis18-lattice-overview}
\end{figure}

The exemplary lattice of SIS18, which is used throughout this contribution, consists of \num{12} sections. An overview is presented in \reff{fig:orm:sis18-lattice-overview}. Each section contains three quadrupoles, labeled \texttt{F}, \texttt{D} and \texttt{T}, and the placement and strength of these quadrupoles is identical in each of the sections. This triplet structure is utilized to increase the transverse acceptance during beam injection. The strength of T-quadrupoles is gradually decreased by one order of magnitude during the ramp, resulting in a small strength during extraction optics.  The \num{36} quadrupoles are connected with five distinct power supplies, separating the quadrupoles into the following families:
\begin{itemize}
    \item \num{6} F-quads from odd numbered sections,
    \item \num{6} F-quads from even numbered sections,
    \item \num{6} D-quads from odd numbered sections,
    \item \num{6} D-quads from even numbered sections,
    \item \num{12} T-quads including all sections.
\end{itemize}

Each section contains two bending magnets next to each other. The horizontal steerers are placed on the first bending magnet, except in sections \num{4} and \num{6} where they are placed on the second bending magnet. The vertical steerers are placed between the F- and the D-quadrupole identically in each of the sections. The vertical and horizontal BPMs are placed downstream of the T-quadrupole, identically in each of the sections.

Each individual electrode of the "shoebox" type capactive pick-up structure is terminated with 50 ohm amplifiers which is followed by direct digitization at 125 MSa/s. The orbit is calculated by least squares fitting the opposite electrode signals on a user defined time window. A detailed discussion on the orbit measurement scheme along with measurement uncertainty estimates can be found in~\cite{SIS18:BPMS:Overview}.

The nominal ORM of SIS18 shows a circulant structure in the vertical block due to the symmetric placement of quadrupoles, vertical steerers and BPMs within each section. In the horizontal block, the circulant structure is broken in the two sections \num{4} and \num{6} because in those sections the horizontal steerer is placed on the second bending magnet rather than the first.

The SIS18 lattice will be used for explaining various important concepts throughout this contribution.

\section{Degeneracy}
\label{sec:degeneracy}

The goal of inverse modeling is to minimize the disagreement between measured and simulated observables. The amount of disagreement is quantified by the \textit{cost function}. Typically, the cost function is given as the "chi-squared" weighted sum of squared deviations:
\begin{equation}
    \chi^2 = \sum_i\frac{\left(m_i - o_i\right)^2}{\sigma_i^2}
\end{equation}
where $o_i$ and $\sigma_i$ are, respectively, the $i$-th observation and measurement uncertainty and $m_i$ is the corresponding simulated quantity obtained from the model. In a more general form, it can be rewritten as
\begin{equation}
    \chi^2 = \bm{r}\bm{\Sigma}\bm{r}
\end{equation}
where $\bm{r} = \bm{m} - \bm{o}$ is the vector of \textit{residuals} and $\bm{\Sigma}$ is the covariance matrix of observations $\bm{o}$.

Any procedure with the goal of predicting a set of model parameters $P$ which minimizes this cost function is referred to as an \textit{estimator}.
The efficiency of an estimator can be quantified by the spread of its predictions around the true parameter values. Thus, the mean squared error ($\mathrm{mse}$) criterion serves as a measure for estimator efficiency:
\begin{equation}\label{eq:degeneracy:estimator-efficiency}
\mathrm{mse}(P) = E\left[(P - \theta)^2\right] = \mathrm{var}(P) + (E\left[P\right] - \theta)^2
\end{equation}
Here, $P$ denotes the predicted parameter values by the estimator, $\theta$ are the true parameter values and $E[\cdot]$ and $\mathrm{var}(\cdot)$ denote, respectively, the expectation value and the variance of its argument. The second term in \refe{eq:degeneracy:estimator-efficiency} corresponds to the bias of the estimator. Thus, regarding the efficiency of an estimator, there is a trade-off between its variance and bias and an increase of the estimator's bias might result in an overall more efficient estimator (reducing the mean squared error of its predictions).

The first mention of quasi-degeneracy for LOCO-like inverse modeling was made in \cite{LOCO:2005:HuangThesis}. The proposed solution was to switch from an unbiased to a biased estimator in order to improve the overall efficiency of the estimates. This was done by augmenting the cost function with terms that correspond to the various specific quasi-degeneracy patterns of the lattice parameters. A related approach~\cite{LOCO:2009:ImprovedFitting} limited the change of lattice parameters during each iteration of the optimization by using a dedicated set of weights in the cost function.

\begin{figure}[hbt]
   \centering
   \includegraphics[width=0.99\textwidth]{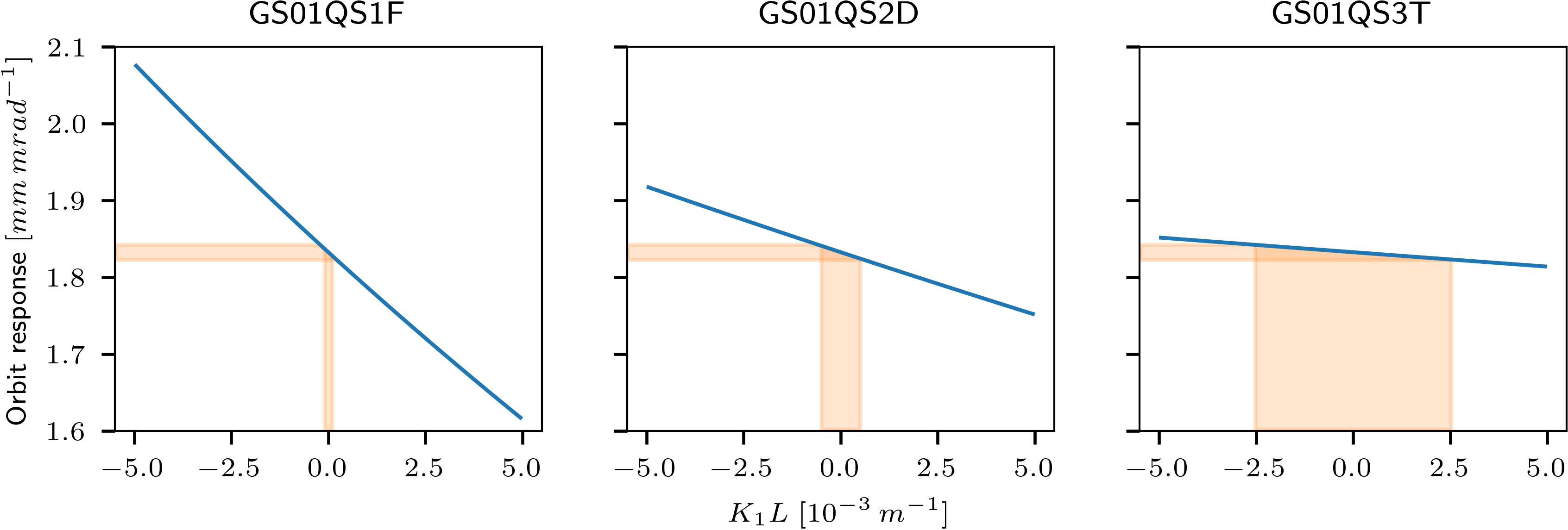}
   \caption{Example for the orbit response change of a single BPM/steerer pair when varying a single quadrupole. The horizontal orange area indicates an orbit response uncertainty of \SI{10}{\micro\metre\per\milli\radian} and is the same for all quadrupoles. The vertical orange area indicates the corresponding plausible region of the quadrupoles' $K_1L$ strengths. Clearly, the  plausible $K_1L$ region is different for the various quadrupoles and it depends on the steepness of the orbit response change with $K_1L$ for each quadrupole.}
   \label{fig:degeneracy:example-one-quadrupole}
\end{figure}

Regarding the terminology, we distinguish between (pure) degeneracy and quasi-degeneracy. A purely degenerate case is one for which there exist multiple distinct solutions that yield the same values for the chosen set of observables in the absence of measurement uncertainty. This is the case if, for example, there are too few BPMs available compared to the number of quadrupoles. A quasi-degenerate case, on the other hand, is one where there exist multiple solutions that are \textit{plausible} in view of the measurement uncertainty, i.e. which can be plausibly explained by the measured data, and some (combinations of) parameters are noticeably more susceptible to the effect of measurement uncertainty than others. The presence of measurement uncertainty doesn't change the nature of the optimization problem though, as there is still a unique global minimum, depending on the specific data used for fitting. Rather, the quasi-degeneracy is a property of the modeled system.
Depending on the lattice and optics, some directions in parameter space will be more "flat" than others and thus are more susceptible to measurement uncertainty. This is sketched in \reff{fig:degeneracy:example-one-quadrupole} where the orbit response of a single BPM/steerer pair is shown in dependence on the three different types of quadrupoles of SIS18, F-, D- and T-quadrupoles. Clearly, the change in orbit response is more flat for the T-quadrupole than for the other two. This example shows only a single ORM element, so for the actual optimization problem the situation is more complex but the principle is the same: flat directions in the parameter space are more susceptible to measurement uncertainty. These directions are determined by the underlying model, i.e. the lattice and optics.

\subsection{Analytical derivative of orbit response}

In order to explain the degeneracy properties for a given lattice, we consider the the orbit response formula $r_{bs}$ for a single dipolar kick and calculate the derivative $r_{kbs}$ with respect to a change in the $k$-th quadrupole's strength.
\begin{equation}
    r_{bs} = \frac{\sqrt{\beta_s\beta_b}}{2\sin(\pi Q)}\cos(\pi Q - |\mu_s - \mu_b|)
\end{equation}
where $b,s$ indicate, respectively, the BPM and steerer index. Taking the derivative with respect to the integrated strength $(K_1L)_k$ of the $k$-th quadrupole, we obtain:
\begin{equation}
\label{eq:quasi-degeneracy:orbit-response-derivative}
\begin{aligned}
r_{kbs} \equiv \frac{dr_{bs}}{d(K_1L)_k} = -r_{bs}\frac{\beta_k}{2} \Bigg\{ & \frac{1}{2\tan(\pi Q)} + \frac{\tan(\pi Q - |\mu_b - \mu_s|)}{2} \\
& + \frac{\cos(2\pi Q - 2|\mu_b - \mu_k|) + \cos(2\pi Q - 2|\mu_s - \mu_k|)}{2\sin(2\pi Q)} \\
& - \frac{\tan(\pi Q - |\mu_b - \mu_s|)}{\sin(2\pi Q)}\int_{\min(\mu_b,\mu_s)}^{\max(\mu_b,\mu_s)}\cos(2\pi Q - |\mu_k - u|)du \Bigg\}
\end{aligned}
\end{equation}
where $\beta_k, \mu_k$ are, respectively, the beta function and phase advance at the $k$-th quadrupole. The full derivation is given in appendix~\ref{seq:appendix:orbit-response-derivative}.

\subsection{Pure degeneracy}

A pure degeneracy exists if there is a set of quadrupoles which can assume different strengths and this is not reflected in the selected observables. Using the ORM as observable this is the case if there are specific lattice segments of quadrupoles without BPMs nor steerers in between.
By considering \refe{eq:quasi-degeneracy:orbit-response-derivative} together with the solution for the integral term given by \refe{eq:appendix:integral-term-explicit}, one can expand the various cosine terms which contain $\mu_k$ contributions by using the trigonometric identity $\cos(x \pm y) = \cos(x)\cos(y) \mp \sin(x)\sin(y)$. For the Jacobian elements corresponding to cases $\mu_b, \mu_s < \mu_k$ (labeled (A)) or $\mu_k < \mu_b, \mu_s$ (labeled (C)), both the cosine terms and the integral term expand into $\sin(2\mu_k)$ and $\cos(2\mu_k)$ terms. For the third case $\mu_{b,s} < \mu_k < \mu_{s,b}$ (labeled (B)), the cosine terms still expand into $\sin(2\mu_k), \cos(2\mu_k)$ while the integral term expands into $\sin(\mu_k)^2, \cos(\mu_k)^2, \sin(\mu_k)\cos(\mu_k)$ terms. By using the trigonometric identities $\sin(2\mu_k) = 2\sin(\mu_k)\cos(\mu_k)$ and $\cos(2\mu_k) = \cos(\mu_k)^2 - \sin(\mu_k)^2$, as well as the trigonometric identity $1 = \cos(\mu_k)^2 + \sin(\mu_k)^2$ for the terms that are independent of $\mu_k$, one can rewrite the whole \refe{eq:quasi-degeneracy:orbit-response-derivative} in terms of $\sin(\mu_k)^2, \cos(\mu_k)^2, \sin(\mu_k)\cos(\mu_k)$ where the coefficients for these terms only depend on $\mu_b$, $\mu_s$ and $Q$. We do not spell out this expanded form of the Jacobian here because it's lengthy and it varies across the three distinct cases (A, B, C). However, an overview of the grouped coefficients is given in the appendix (\reft{tab:appendix:proof:coefficient-vector-expressions}). In the following, we focus on the following more general observations. Given that the Jacobian for each BPM/steerer/quadrupole triple can be written as the sum of three expressions involving $\mu_k$ (namely, $\sin(\mu_k)^2, \cos(\mu_k)^2, \sin(\mu_k)\cos(\mu_k)$) together with their coefficients which depend solely on $\mu_b$, $\mu_s$, $Q$, each column of the Jacobian can be written as a linear combination of $\bm{v}_1\sin(\mu_k)^2 + \bm{v}_2\cos(\mu_k)^2 + \bm{v}_3\sin(\mu_k)\cos(\mu_k)$ where the column vectors $\bm{v}_{1,2,3}$ contain the row-wise constant coefficients depending only on $\mu_b$, $\mu_s$, $Q$. The expressions for these coefficients are the same for each group of quadrupoles that is not interleaved by BPMs nor steerers. Thus, the column span of the Jacobian is given by the three column vectors $\bm{v}_{1,2,3}$ for each group of quadrupoles and thus, for a lattice with $N$ sections and \num{3} or more non-interleaved quadrupoles per section, the rank of the Jacobian is at most $3N$. It should be emphasized that this holds only if all the involved quadrupoles in each section are consecutive, i.e. not interleaved by BPMs nor steerers, since otherwise their coefficients would change according to the cases (A, B, C). This implies that 4 or more consecutive quadrupoles per section will cause a pure degeneracy since their contributions to the Jacobian can still be described by only three column vectors. This result holds for one dimension (horizontal or vertical) but for uncoupled optics it is easily extended to both dimensions by considering that there are $\sin(\mu_k)^2, \cos(\mu_k)^2, \sin(\mu_k)\cos(\mu_k)$ terms for both dimensions separately, i.e. six independent coefficient vectors $\bm{v}_{1,2,3,4,5,6}$. Thus, the dimension of the column span of the Jacobian involving both dimensions is bounded by $6N$ and, therefore, 7 or more consecutive quadrupoles will cause a pure degeneracy.

\begin{figure}[hbt]
   \centering
   \includegraphics[width=.49\textwidth]{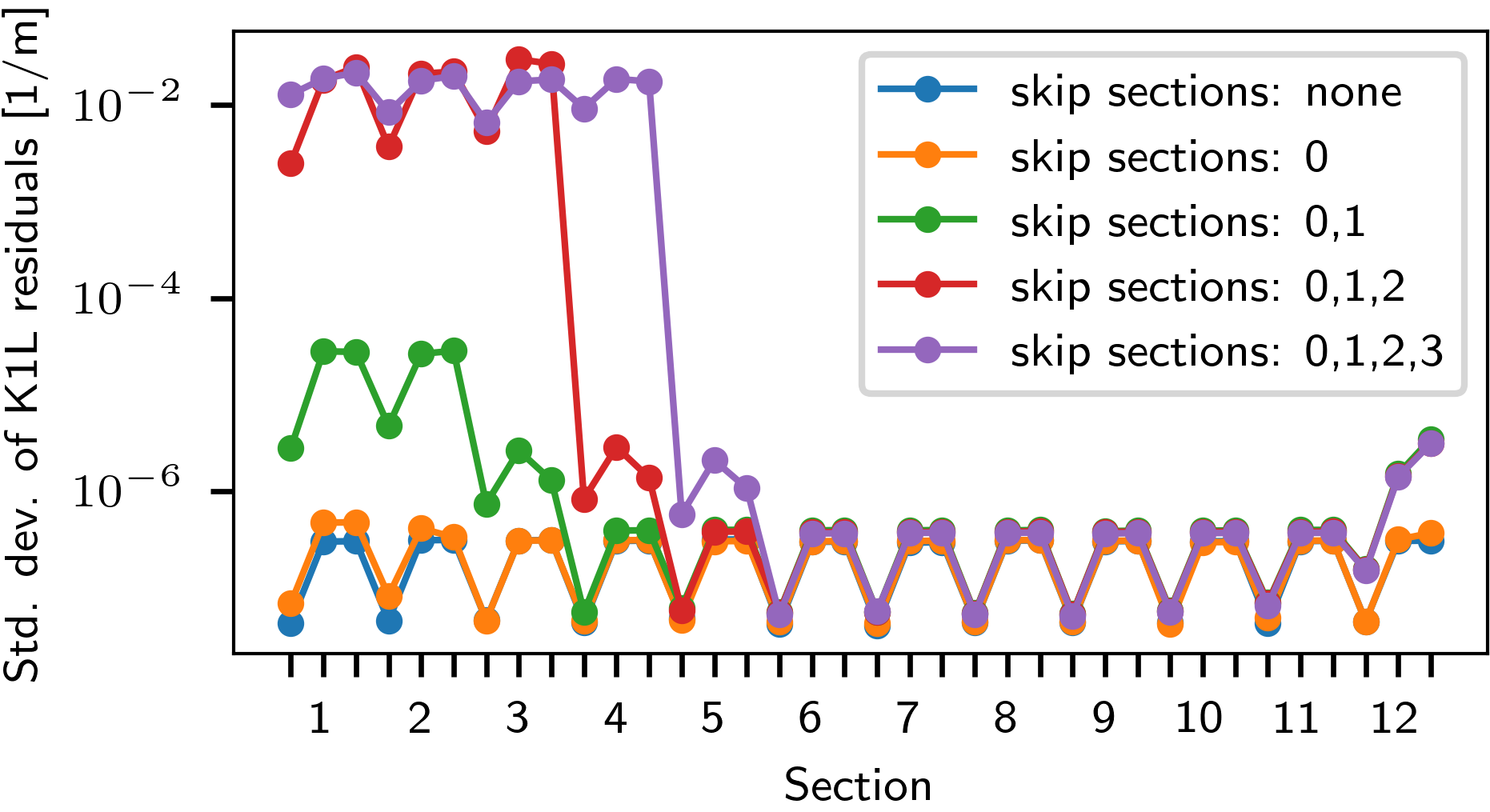}
   \caption{$K_1L$ residuals when running Levenberg-Marquardt optimization for the nominal optics, starting from \SI{1}{\percent} random quadrupole errors, and gradually leaving out BPMs and steerers from consecutive sections in order to cause a pure degeneracy of the inverse problem. Each tick marker on the horizontal axis indicates a quadrupole (F-, D-, T-quadrupole per section).}
   \label{fig:degeneracy:pure-degeneracy}
\end{figure}

This is in agreement with the result derived in \cite{LocalOpticsCorrection} which is that for uncoupled transverse optics, a set of 7 or more consecutive quadrupoles in both dimensions (or 4 or more quadrupoles in one dimension) can produce locally confined optics variations in between their segment. Since the orbit response is a specific combination of the lattice optics, and it depends only on the optics at the BPM and steerer locations as well as the tune, if there exist such segments of quadrupoles not interleaved with BPMs nor steerers, the optics within such segments cannot be resolved by observing the ORM. This can be seen from \reff{fig:degeneracy:pure-degeneracy} which shows simulated inverse modeling results for the SIS18 lattice for all 36 quadrupoles, without any simulated measurement uncertainty, while leaving out the BPMs and steerers from an increasing number of consecutive sections. As can be seen, for the cases where none of the sections or only the first section is skipped, the quadrupole strengths can be reliably recovered down to the numerical precision of the estimator. When three or four consecutive sections are skipped, the estimates clearly become ambiguous which is reflected by the large increase in their standard deviation. This is because each section contains three distinct quadrupoles and hence, when skipping three or more sections, the corresponding segment contains more than \num{7} quadrupoles required to exhibit a degeneracy. For the case where two sections are skipped, i.e. six quadrupoles, there is a slight increase in standard deviation, similar to the amount that's visible for the neighboring sections in the skip-3 and skip-4 cases. This is because when the degenerate segment is extended with its neighboring sections, the variations induced by those quadrupoles at the boundaries of the segment are on the level of the numerical precision of the estimator and hence won't be distinguished. Nevertheless, it should be noted that the order of magnitude is much smaller.
The saw-tooth pattern that can observed between D- and T-quadrupoles will be explained as quasi-degeneracy below.

\subsubsection{Global degeneracy}
\label{sec:quasi-degeneracy:global-degeneracy}

Besides the intra-section degeneracy discussed above, which is caused by isolated groups of consecutive quadrupoles, there can be another, global degeneracy whose existence also depends on the BPM/steerer placement. In the following, we use the notation \texttt{S,Qn+,B} which means that we are considering one dimension (horizontal or vertical) and the placement of lattice elements within a section is the following: steerer, followed by $n$ quadrupoles (\texttt{n+} means $n$ or more), followed by a BPM. In terms of the results this is similar to \texttt{B,Qn+,S}. This pattern describes the placement for one section and is repeated on a section-to-section basis. We emphasize that this only describes in what order BPM, steerer and quadrupoles are placed but it doesn't restrict the specific locations in terms of the phase advance within each section. In fact, these specific locations may be different from section to section. For both dimensions, horizontal and vertical, we write \texttt{Sh,Sv,Qn+,Bh,Bv}, where \texttt{h} refers to horizontal and \texttt{v} refers to vertical. In terms of the results this is similar to any other pattern that swaps any steerer with any BPM. This is because the Jacobian only depends on $|\mu_b - \mu_s|$ and it separates horizontal from vertical contributions.

We show that the following placements exhibit a global degeneracy: \texttt{S,Q3+,B} and \texttt{Sh,Sv,Q5+,Bh,Bv}. It's worth noting that \texttt{Sh,Sv,Q5,Bh,Bv} causes a rank deficiency of degree 1 in the Jacobian while \texttt{Sh,Sv,Q6,Bh,Bv} causes a degree 2 rank deficiency. For \texttt{Sh,Sv,Q7+,Bh,Bv} intra-section degeneracy will appear and the rank of the Jacobian is the same as for \texttt{Sh,Sv,Q6,Bh,Bv}. The argument for this is similar to the one for \texttt{S,Q4+,B} above, since exactly three column vectors are needed for each dimension in order to generate the Jacobian columns for a group of consecutive quadrupoles in that dimension. In the appendix, we proof the rank deficiency for the \texttt{S,Q3+,B} (appendix~\ref{sec:appendix:proof:sqqqb}) and \texttt{Sh,Sv,Q6+,Bh,Bv} (appendix~\ref{sec:appendix:proof:ssq6bb}) placements. The origin of the rank deficiency for the \texttt{Sh,Sv,Q5,Bh,Bv} pattern is not obvious and we report this without proof, based on our simulation results. Table~\ref{tab:quasi-degeneracy:overview-of-global-degeneracies} gives an overview of the various Jacobians' ranks obtained via simulations, in agreement with the analytical derivations.

\begin{table}[hbt]
\centering
\caption{This table presents an overview of the Jacobian properties in terms of rank deficiency for the various BPM/steerer placements around groups of consecutive quadrupoles. $N$ denotes the number of sections in the lattice ($N\geq 3$ is assumed). It should be emphasized that the only deciding factor is the placement pattern, i.e. how many quadrupoles form a consecutive group, not where exactly these quadrupoles or the BPMs/steerers are located in each of the sections. The specific locations may vary from section to section and as long as the overall placement pattern is satisfied, the rank deficiency will be the same. The Jacobians were obtained from simulations using the mpmath~\cite{mpmath} library to avoid numerical issues (\texttt{dps} set to \num{100}). The rank is then computed as the number of singular values that are larger than or equal to $\epsilon N^2 s_{\max}$ where $s_{\max}$ is the largest singular value and $\epsilon = 2^{-52}$ is the machine epsilon for double precision floating point numbers.}
\begin{tabular}{|cccc|}
\hline
& \multicolumn{3}{c|}{Jacobian} \\
& \# rows & \# columns & rank \\
\hline
\texttt{S,Q2,B} & $N^2$ & $2N$ & $2N$ \\
\texttt{S,Q3,B} & $N^2$ & $3N$ & $3N-1$ \\
\texttt{S,Q4+,B} & $N^2$ & $4^{\small +}N$ & $3N-1$ \\
\hline
\texttt{Sh,Sv,Q4,Bh,Bv} & $2N^2$ & $4N$ & $4N$ \\
\texttt{Sh,Sv,Q5,Bh,Bv} & $2N^2$ & $5N$ & $5N-1$ \\
\texttt{Sh,Sv,Q6,Bh,Bv} & $2N^2$ & $6N$ & $6N-2$ \\
\texttt{Sh,Sv,Q7+,Bh,Bv} & $2N^2$ & $7^{\small +}N$ & $6N-2$ \\
\hline
\end{tabular} 
\label{tab:quasi-degeneracy:overview-of-global-degeneracies}
\end{table}

The appendix~\ref{sec:appendix:orbit-response-derivative-beamlines} includes a similar derivation for beamlines, i.e. non-circular lattices.

\subsection{Quasi-degeneracy}

Even though groups of, for example, two consecutive quadrupoles do not exhibit a pure degeneracy, they can exhibit a quasi-degeneracy which means that their estimated strengths are much more susceptible to measurement uncertainty than the ones of other quadrupoles. This type of quasi-degeneracy is explained in the following section.

The covariance of parameter estimates under linear least squares is given by $\sigma^2(J^T J)^{-1}$ where $\sigma^2$ is the variance of observables and $J$ is the Jacobian (if errors are heteroscedastic, it is $(J^T \Sigma J)^{-1}$ with $\Sigma$ being the covariance matrix of observables).
This is closely related to the matrix $J^T J$. The eigenvectors of a matrix and its inverse are similar and the eigenvalues are reciprocal, so studying the matrix $J^T J$ reveals important information about the error propagation.
Also, in Gauss-Newton minimization, $J^T J$ is used as an approximation of the Hessian $H$ and thus, a lower bound for the estimated parameter variance is given by $\sigma^2 H^{-1}$. This is, of course, in agreement since at the minimum of the cost function, the gradient is assumed to vanish, so the flatness of the cost function depends on how quickly that zero gradient changes in the neighborhood of the estimate which is indicated by the Hessian matrix.

\begin{figure}[hbt]
   \centering
   \includegraphics[width=0.99\textwidth]{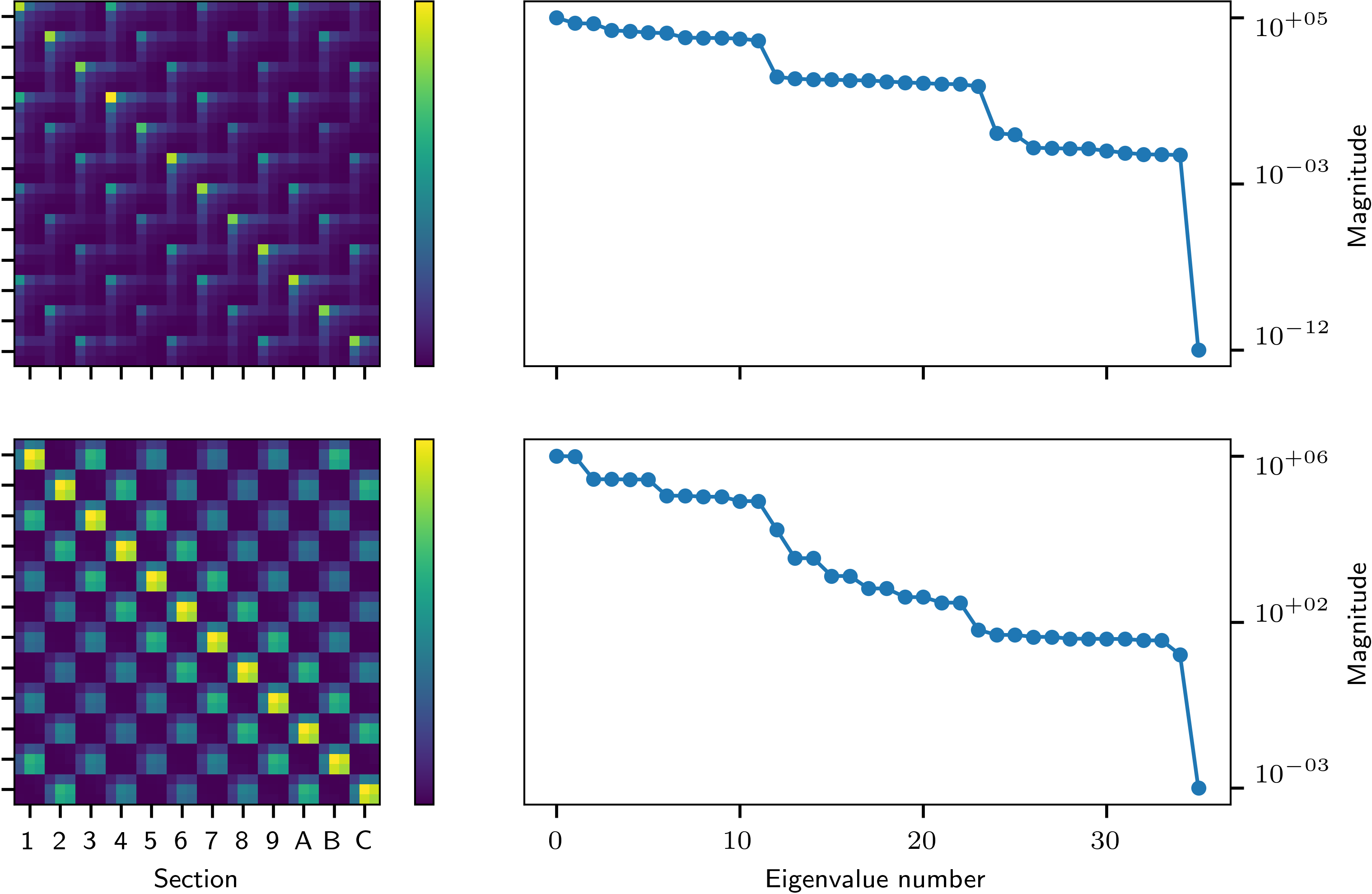}
   \caption{Top row: Horizontal dimension, bottom row: vertical dimension. Left column: The $36\times 36$ matrix $J^T J$. The axes numbering indicates the \num{12} sections of SIS18 in hexadecimal notation and there are 3 rows/columns per section, corresponding to the F-, D- and T-quadrupoles (in that order) of each section. Right column: The eigenvalues of corresponding $J^T J$ matrices. The color bars and eigenvalue magnitude indicate the magnitude of $J^T J$ in units of $\si{\metre^4/\radian^2}$. The values of the color bar correspond to those of the eigenvalue plots shown on the vertical axes. For the horizontal dimension, the smallest eigenvalue $\lambda_{35}$ is nonzero only due to limited floating point precision. When inspecting the \num{12} smallest horizontal eigenvalues, it can be observed that the $\lambda_{24},\lambda_{25}$ eigenvalues have a slightly greater magnitude than the remaining \num{9} eigenvalues (neglecting $\lambda_{35}$). These two eigenvalues correspond to the sections 4 and 6 where the horizontal steerer is shifted by a few meters compared to the other sections.}
   \label{fig:degeneracy:jj-sep-dims}
\end{figure}

Figure~\ref{fig:degeneracy:jj-sep-dims} shows the $J^T J$ matrices emerging from horizontal and vertical ORMs, together with their eigenvalue spectra.  There are a few things to be noted. First of all, for the vertical $J^T J$ plot it can be seen that it indicates higher variance for the D-T-quadrupole pairs than for the F-D- or F-T-pairs. This is because of the scaling of the Jacobian with the beta function which, in vertical, is larger at D- and T-quadrupoles than at F-quadrupoles (see \reff{fig:orm:sis18-lattice-overview}). Secondly, it can be observed that in both dimensions there is one eigenvalue that is much smaller than others. Small eigenvalues of $J^T J$ correspond to large eigenvalues of $(J^T J)^{-1}$, i.e. of the covariance estimate for model parameters. However, for the horizontal $J^T J$ matrix, since in horizontal lattice features a \texttt{S,Q3,B} steerer/BPM placement which causes a pure degeneracy (see section~\ref{sec:quasi-degeneracy:global-degeneracy}), the smallest eigenvalue in this plot is only nonzero due to limited numerical precision. A zero eigenvalue for $J^T J$ implies a pure degeneracy since the system $J^T J\Delta p = J^T r$ ($\Delta p$ parameter update, $r$ residuals) is under-determined. That is, the null space of $J^T J$ is nonzero and, hence, there exists a parameter update $\Delta p$ that will leave the residuals unchanged at zero.
In general, a small eigenvalue for $J^T J$ implies a direction of quasi-degeneracy which is given by the corresponding eigenvector. It means that the parameter update emerging from $J^T J\Delta p = J^T r$ will be susceptible to measurement uncertainty in the direction of the corresponding eigenvector. This is what is observed for the vertical Jacobian where the vertical lattice features a \texttt{S,Q2,B} BPM/steerer placement.

\begin{figure}[hbt]
   \centering
   \includegraphics[width=0.49\textwidth]{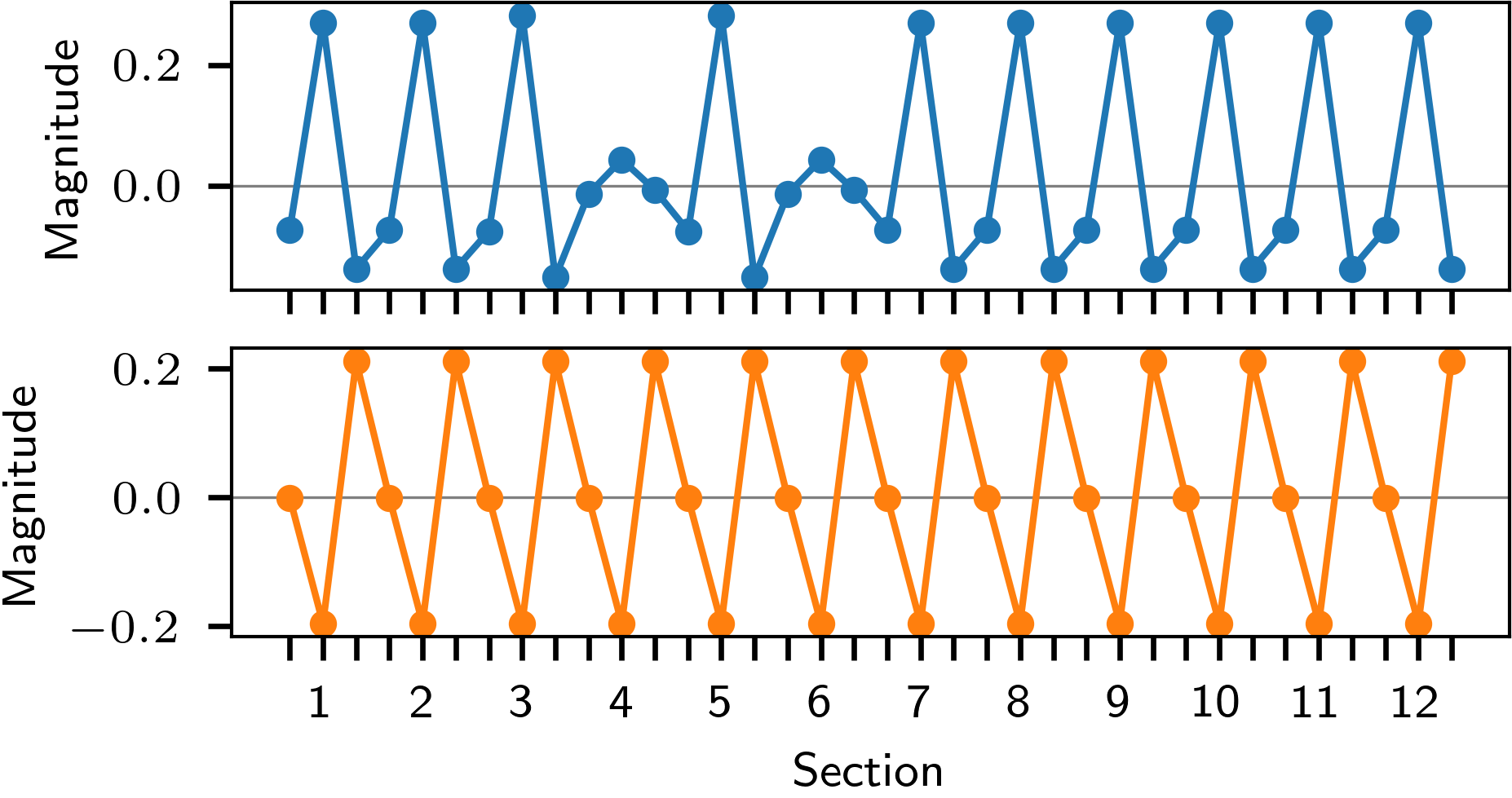}
   \caption{Eigenvectors that corresponds to the smallest eigenvalue of the $J^T J$ matrices in horizontal (top) and vertical (bottom) dimension. Each tick marker on the horizontal axis indicates a quadrupole (F-, D-, T-quadrupole per section).}
   \label{fig:degeneracy:jj-smallest-eigenvector}
\end{figure}

Figure~\ref{fig:degeneracy:jj-smallest-eigenvector} shows the two eigenvectors, in horizontal and vertical dimension, that correspond to the smallest eigenvalue of the corresponding $J^T J$ matrix. Since the eigenvectors of a matrix and its inverse are similar, these indicate the direction of (quasi-)degeneracy in both dimensions separately. It can be observed that this is a global degeneracy in both cases, since all quadrupoles participate; hence, there is only one eigenvalue that is significantly smaller than all others. This is due to the symmetry of the lattice with respect to the BPM/steerer placement pattern. In horizontal, for the two sections 4 and 6 where the ORM's circulant structure is broken, it can be observed that a corresponding change in the quadrupole's degeneracy pattern reflects this. In vertical, it can be observed that the quasi-degeneracy is driven by the (non-interleaved) D-T-quadrupole pairs.

\begin{figure}[hbt]
   \centering
   \includegraphics[width=0.99\textwidth]{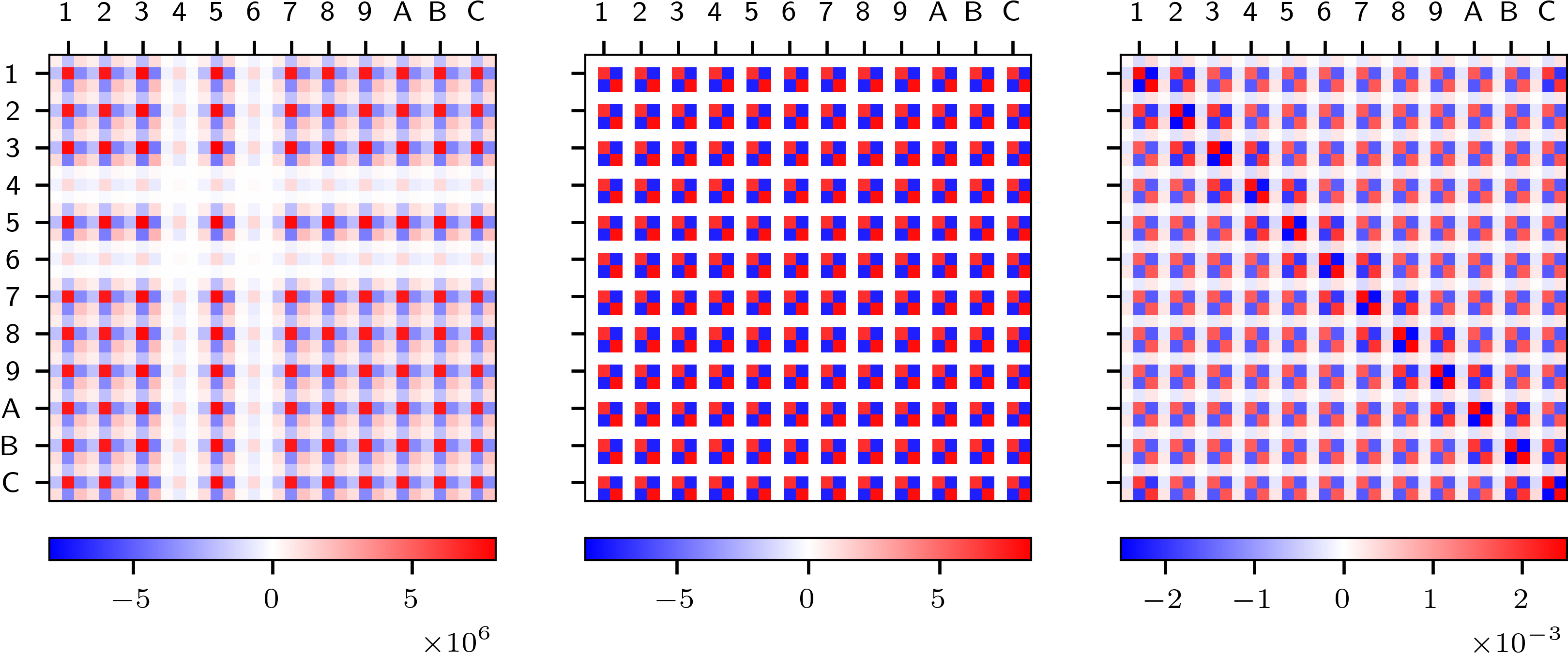}
   \caption{$(J^T J + \alpha I)^{-1}$ for the horizontal ($\alpha=\num{1d-8}$; left), vertical ($\alpha=0$; middle) and combined ($\alpha=0$; right) dimensions. The axes numbering indicates the \num{12} sections of SIS18 in hexadecimal notation and there are 3 rows/columns per section, corresponding to the F-, D- and T-quadrupoles (in that order) of each section. The unit of the color bars, indicating the magnitude of the matrices, is \si{\radian^2/\metre^4}. The quasi-degeneracy pattern looks very symmetric in vertical because the BPM/steerer placement is fully symmetric from section to section. This, however, is not a requirement as shown by the horizontal data. The degeneracy pattern reflects the differently placed steerers in section 4 and 6. In fact, no symmetry whatsoever in terms of the exact phase advances of BPMs or steerers is required for a degeneracy pattern to occur; only the placement pattern in terms of upstream or downstream of quadrupoles is deciding. For the combined dimensions it can be observed that the resulting pattern is not fully symmetric but features local correlations slightly more than ones with other sections. This is because the magnitude of the smallest eigenvalue for the combined dimensions is closer to the magnitude of other eigenvalues and thus does not dominate the pattern alone.}
   \label{fig:degeneracy:jj-inv-sep-dims}
\end{figure}

Figure~\ref{fig:degeneracy:jj-inv-sep-dims} shows the scaling of the covariance estimate for model parameters, i.e. $(J^T J)^{-1}$; for horizontal, since it is rank deficient, $(J^T J + \alpha I)^{-1}$ is plotted (with $\alpha = \num{1d-8}$ i.e. Tikhonov regularized, which is also used by e.g. the Levenberg-Marquardt optimizer, though it uses a flexible regularization parameter $\alpha$). Clearly, the global nature of the degeneracy is reflected in the eigenvectors \reff{fig:degeneracy:jj-smallest-eigenvector}. From \reff{fig:degeneracy:jj-sep-dims} it can be observed that pairwise cancellation is mostly confined to nearby sections and decreases when moving further away in terms of the phase advance. However, the final covariance of quadrupole estimates is dominated by a strong global component which is symmetric for the vertical ORM.

For the vertical ORM, the corresponding $J^T J$ matrix is a block circulant matrix by the argument of section-to-section symmetry of the vertical lattice. The eigenvectors of a block circulant matrix $B = \mathrm{bcirc}(\bm{b}_0,\bm{b}_1,\dots,\bm{b}_{n-1}) \in \mathcal{BC}_{n,k}$ (where $n$ is the number of blocks and $k$ the size of a $k\times k$ block; i.e. $n=12$, $k=3$ in our case) are derived in \cite{Tee:BlockCirculantMatrices}. They are given by:
\begin{equation}\label{eq:quasi-degeneracy:block-circulant-eigenvectors}
\begin{bmatrix}
    \bm{v} \\
    \rho_m\bm{v} \\
    \rho_m^2\bm{v} \\
    \vdots \\
    \rho_m^{n-1}\bm{v} \\
\end{bmatrix}
\end{equation}

where $\bm{v}$ is a nonzero column vector of length $k$, which is given below, and $\rho_m$ is one of the $n$ complex roots of unity: $\rho_m = \exp(2\pi i \frac{m}{n})$. For each $\rho_m$ there are $k$ distinct vectors $\bm{v}$ given by the eigenvector equation~\cite{Tee:BlockCirculantMatrices}:
\begin{equation}
(\bm{b}_0 + \rho\bm{b}_1 + \rho^2\bm{b}_2 + \dots + \rho^{n-1}\bm{b}_{n-1})\bm{v} = \lambda\bm{v}
\end{equation}
where $\lambda$ is the corresponding eigenvalue.

Since the first of the $n$ roots of unity is $\rho_0 = 1$, from \refe{eq:quasi-degeneracy:block-circulant-eigenvectors} it becomes apparent that every block circulant matrix $B\in\mathcal{BC}_{n,k}$ has exactly $k$ distinct globally symmetric eigenmodes which repeat on a block-to-block basis. This is the case for the vertical $J^T J$ matrix.

\begin{figure}[hbt]
   \centering
   \includegraphics[width=0.99\textwidth]{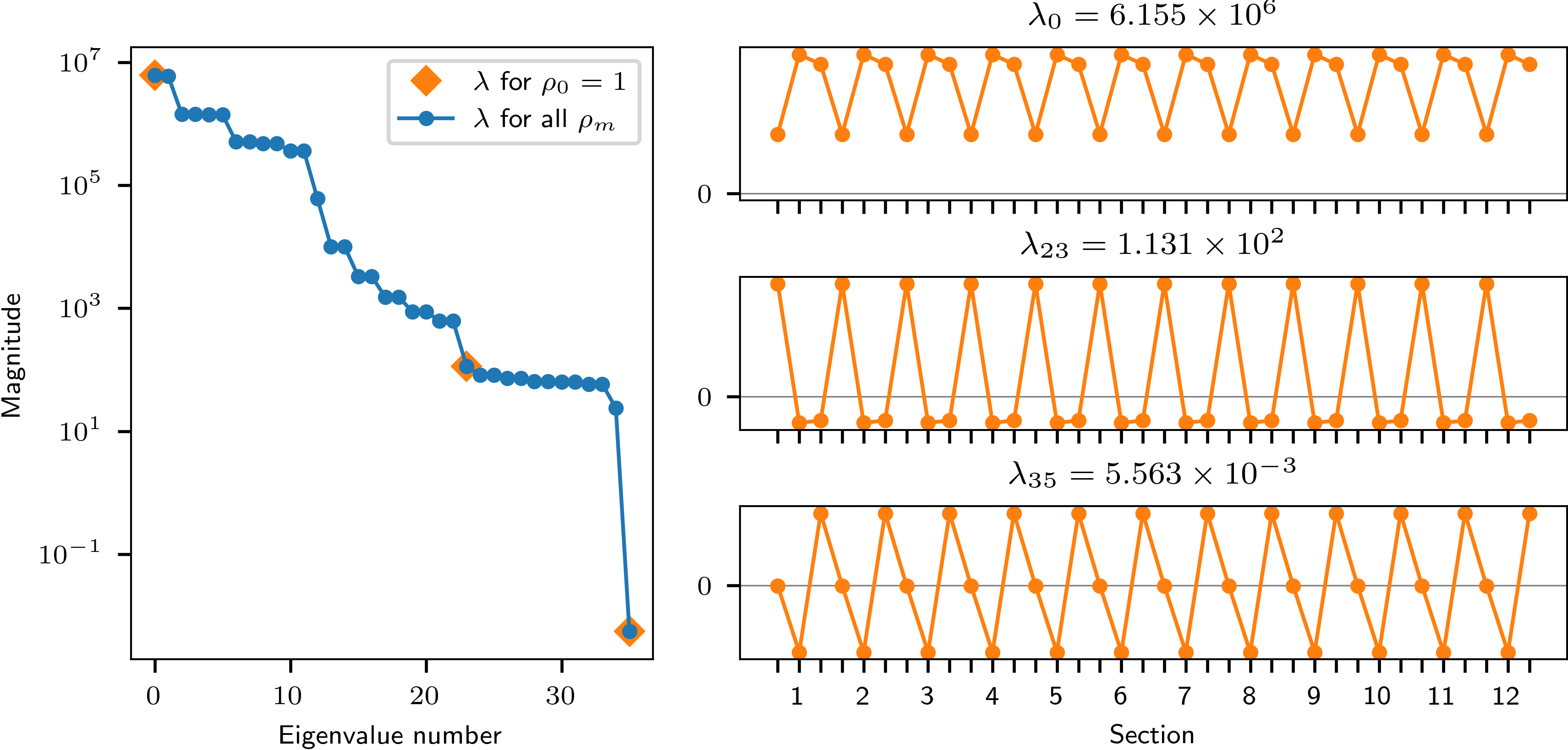}
   \caption{Globally symmetric eigenmodes of $J^T J$ in vertical dimension which arise due to the fact that the vertical lattice is symmetric from section to section. Each tick marker on the horizontal axis of the eigenvector plots indicates a quadrupole (F-, D-, T-quadrupole per section).}
   \label{fig:degeneracy:jj-vertical-global-eigenmodes}
\end{figure}

Because $J^T J$ is real and symmetric, its eigenvalues are guaranteed to be real, too. Furthermore, since $J^T J$ is a Gram matrix, it is positive semidefinite and its eigenvalues are guaranteed to be greater than or equal to zero.
This is observed for the vertical $J^T J$ matrix and it happens that one of the globally symmetric eigenmodes is associated with the smallest eigenvalue $\lambda_{35}$. Figure~\ref{fig:degeneracy:jj-vertical-global-eigenmodes} shows the three globally symmetric eigenmodes corresponding to the $\rho_0 = 1$ eigenvalues.

Because for the horizontal lattice, the circulant structure of the ORM and thus of $J^T J$ is broken in the two sections 4 and 6, it can't have a globally symmetric eigenmode, i.e. a mode that repeats on a section-to-section basis. However, as becomes apparent from the eigenvector \reff{fig:degeneracy:jj-smallest-eigenvector}, the global mode still affects all sections at once and reflects the breaking of symmetry in the sections 4 and 6.

\subsection{Example}

In the absence of BPM errors, inverse modeling with an optimizer such as Levenberg-Marquardt will always converge to the ground truth solution (within the boundaries of numerical precision), given that the there is no additional model bias present and the initial guess is not too far from the ground truth (so that the optimizer won't cross any instabilities, for example).

\begin{figure}[hbt]
   \centering
   \includegraphics[width=.49\textwidth]{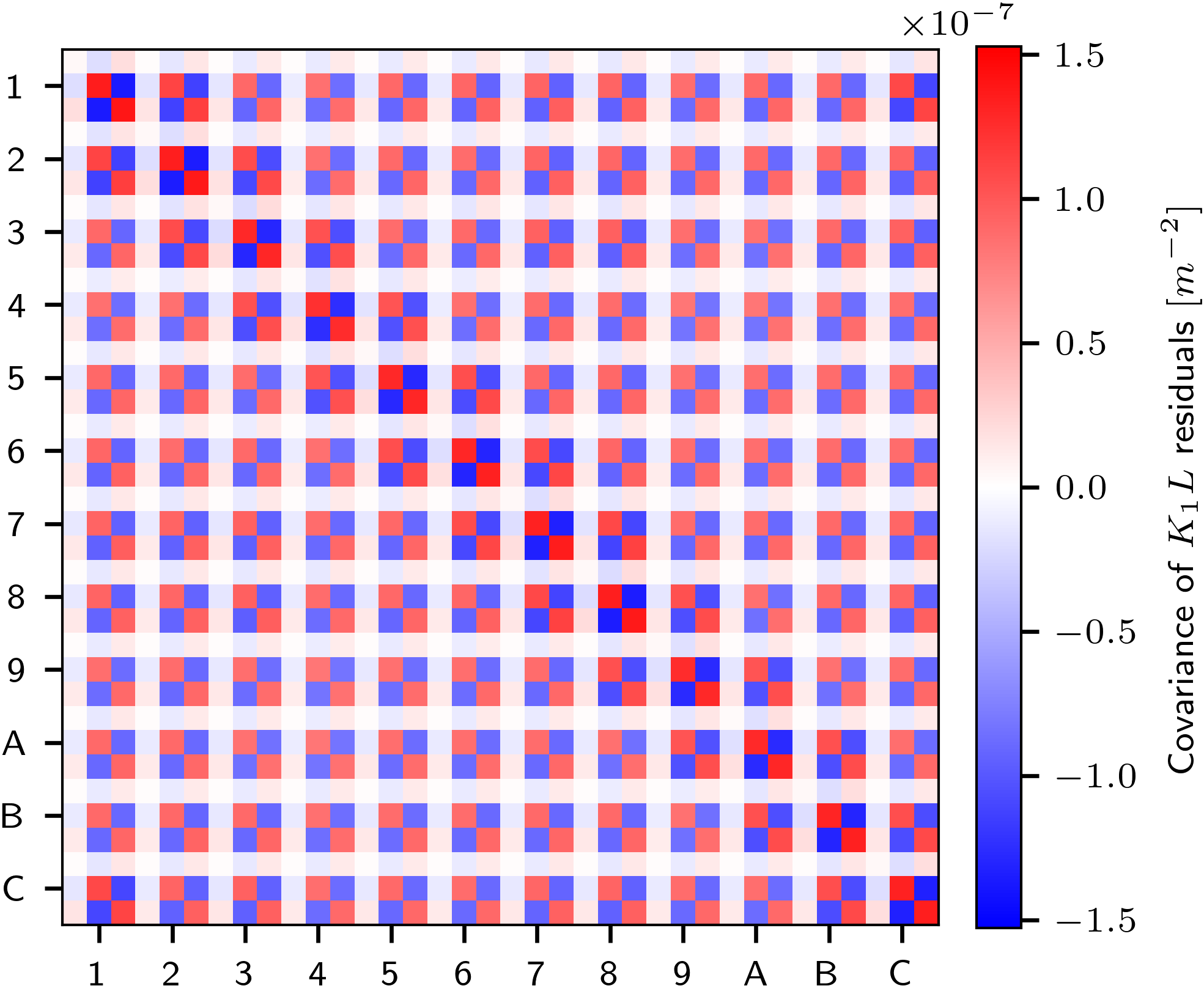}
   \caption{Covariance of $K_1L$ residuals obtained with Levenberg-Marquardt optimizer for \SI{7}{\micro\metre\per\milli\radian} orbit response uncertainty when including both horizontal and vertical ORM. The axes numbering indicates the \num{12} sections of SIS18 in hexadecimal notation and there are 3 rows/columns per section, corresponding to the F-, D- and T-quadrupoles (in that order) of each section. No quadrupole errors were applied to the lattice and optimization started at the nominal quadrupole strengths. Thus, the $K_1L$ residuals emerge purely as a result of the simulated ORM uncertainty. All \num{36} quadrupoles have been included in the optimization.}
   \label{fig:degeneracy:only-bpm-noise-covariance-of-residuals}
\end{figure}

Figure~\ref{fig:degeneracy:only-bpm-noise-covariance-of-residuals} shows the covariance of the various solutions obtained with Levenberg-Marquardt optimizer when no quadrupole errors are applied to the lattice and only BPM errors are present in the ORM simulation. That is, each of the inverse modeling instances is given a distinct noisy ORM emerging from the same orbit response uncertainty of \SI{7}{\micro\metre\per\milli\radian}. The initial guess is the ground truth solution, i.e. no quadrupole errors, but from the perspective of the optimizer this is not the minimum of the cost function due to the noise in the ORM; hence, it will converge to a different solution, the $K_1L$ residuals. The structure of these solutions is determined by the underlying simulation model including the lattice optics.
It can be seen that the quasi-degeneracy is mainly driven by the D-T-quadrupole pairs where much larger excursions in $K_1L$ residuals happen. This is in agreement to \reff{fig:degeneracy:jj-inv-sep-dims} which shows the predicted uncertainty from the Jacobian. For \SI{7}{\micro\metre\per\milli\radian} orbit response uncertainty, the expected covariance of D- and T-quadrupole strengths is approximately $(\num{7d-3})^2\cdot 0.002\,\si{\per\square\metre} \approx \SI{1d-7}{\per\square\metre}$. This is the amount that can be observed from the simulations with Levenberg-Marquardt optimizer in \reff{fig:degeneracy:only-bpm-noise-covariance-of-residuals}. Also, the observed covariance pattern matches with the one from \reff{fig:degeneracy:jj-inv-sep-dims}.

\subsection{Counteracting quasi-degeneracy}

At different stages, different options for counteracting degeneracy are feasible. During the design phase of the accelerator, the placement of steerers and BPMs can be investigated in order to find a placement that reduces the amount of quasi-degeneracy compared to other placement candidates. For the SIS18 lattice, this would be achieved by positioning the BPMs between the D- and T-quadrupoles. At the stage of data analysis, the choice of optimizer allows for different strategies to counteract the quasi-degeneracy. Examples include SVD cutoff or adding additional constraints to the cost function.

\subsubsection{Placement of BPMs/steerers}

\begin{figure}[hbt]
   \centering
   \includegraphics[width=0.49\textwidth]{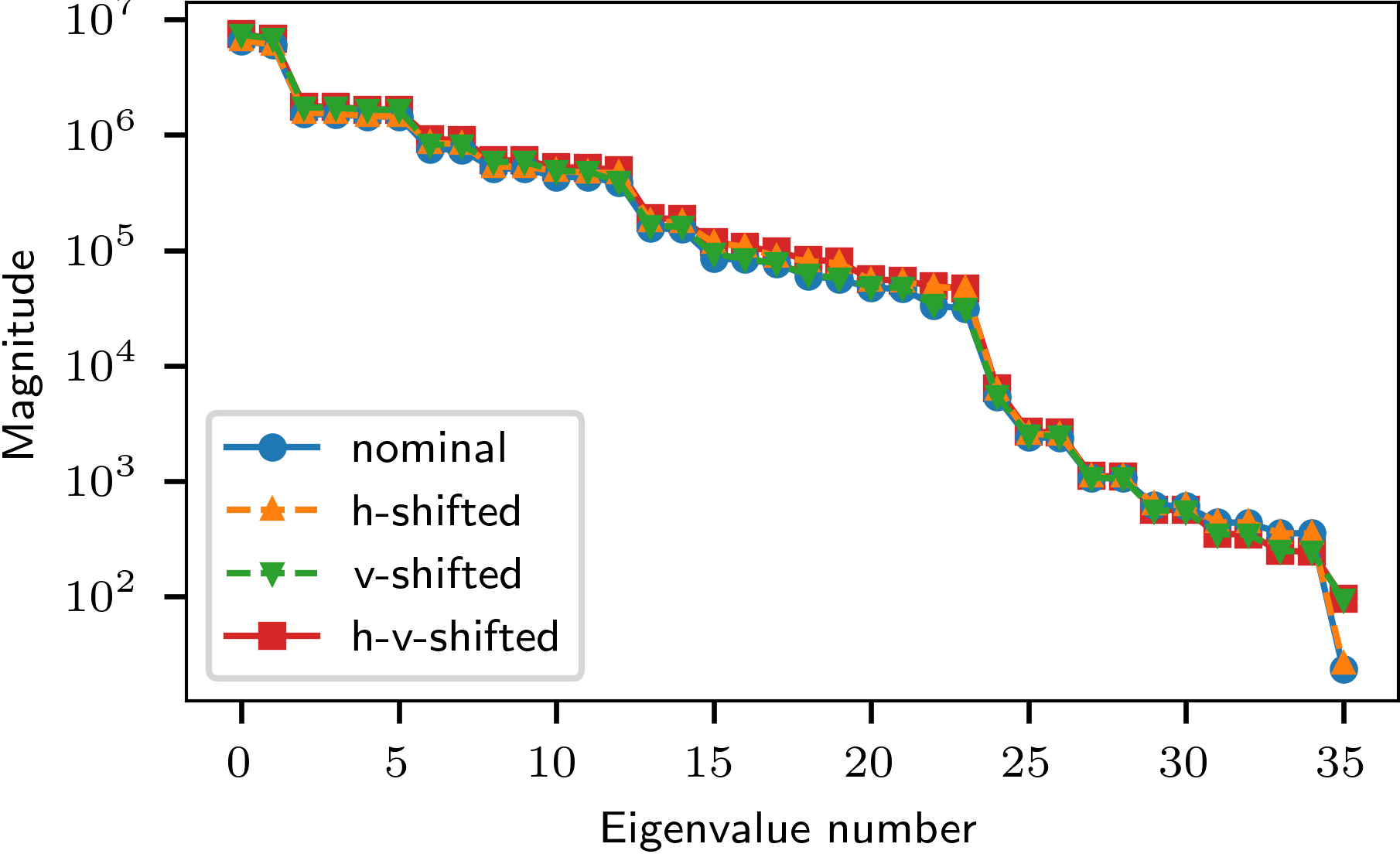}
   \caption{Eigenvalues of $J^T J$ for different BPM placements. \textit{nominal} refers to the original lattice, \textit{h-shifted} refers to the lattice where the horizontal BPM has been shifted from its original position (downstream of the T-quadrupole) to in between the D- and T-quadrupole. \textit{v-shifted} means the same for the vertical BPM and \textit{h-v-shifted} refers to both BPMs being shifted between the D- and T-quadrupole. The different placement strategies vary in their smallest eigenvalue which is the one that drives the propagation of uncertainty.}
   \label{fig:degeneracy:jj-eigenvalues-hv-bpm-shift}
\end{figure}

At a stage where this is still possible, the careful planning of BPM/steerer locations can help to avoid or mitigate quasi-degeneracy.
We compare the following three scenarios with the results for the nominal lattice: moving either the horizontal or vertical BPM or both BPMs between the D- and T-quadrupole. Figure \ref{fig:degeneracy:jj-eigenvalues-hv-bpm-shift} shows the $J^T J$ eigenvalue spectra for these three cases as well as for the nominal case. It can be observed that the different placements of BPMs have different effects on the amount of quasi-degeneracy. Specifically, the versions where the vertical BPMs are shifted between the D- and T-quadrupole yields significantly smaller uncertainty in the estimated parameters while the version with only horizontal BPMs shifted has a negligible effect. Thus, it is of importance to explore the different options for BPM placement in order to allow for more precise inverse modeling results for future accelerators.

\section{Fitting of the orbit response matrix}
\label{sec:fitting-with-jacobian}

The Levenberg-Marquardt optimizer uses the Jacobian at every iteration. Typically, this Jacobian is computed numerically via finite-difference approximation, with an appropriate step size $\Delta$ for each parameter. In the following, we use the analytically derived Jacobian (see~\refe{eq:quasi-degeneracy:orbit-response-derivative}), which is obtained from Twiss data, for the optimization procedure. While there is a mismatch between the numerical (real) and the analytical Jacobian, if this mismatch is manageable then the fitting will still converge. This has similarities to how closed orbit feedback (COFB) correction with model mismatch works~\cite{COFB:spatial-model-mismatch}. In the context of COFB, the system is assumed to be linear and there exists a true response matrix $R$ and a model response matrix $R_{\Theta}$. In an iterative scheme, the COFB converges if the eigenvalues $\lambda_i$ of $1 - RR_{\Theta}^{+}$ fulfill $-1 \leq \lambda_i \leq 1$ (where the superscript ${}^+$ denotes the pseudo-inverse). If $R$ and $R_{\Theta}$ are square matrices, the relationship has to be a strict inequality to achieve convergence, i.e. $-1 < \lambda_i < 1$. Otherwise, if $R$ and $R_{\Theta}$ are $m\times n$ matrices with $m > n$, then $1 - RR_{\Theta}^{+}$ must have largest eigenvalue $1$ with multiplicity $m-n$ and all other eigenvalues must fulfill $-1 < \lambda_i < 1$.
In the context of LOCO fitting, the matrices $R$ and $R_{\Theta}$ denote, respectively, the true and analytical Jacobian. Also, the system is not entirely linear, so the lattice model reacts differently to a parameter update than the linear transformation given by $R$. However, if the magnitude of updates is constrained, a locally linear version can be assumed at every iteration. This implies a varying true matrix $R \equiv R(x)$ where $x$ is the current guess of model parameters. For an iterative scheme to converge, the eigenvalues of the sequence of matrix multiplications
\begin{equation}\label{eq:fitting:feedback-matrix-mult-sequence}
\left(1 - RR_{\Theta}^{+}\right)_{k-1} \; \dots \; \left(1 - RR_{\Theta}^{+}\right)_{0}
\end{equation}
must tend to zero as $k\rightarrow\infty$ (where $k$ denotes the iteration count; except the $m-n$ excess eigenvalues for rectangular $R,R_{\Theta}$ remain at $1$). This is provided if the eigenvalues of the individual matrices $(1 - RR_{\Theta}^{+})_{i}$, for guess $x_i$ during each iteration, fulfill $-1 \leq \lambda_i \leq 1$, i.e. if the model mismatch is manageable for each relevant optics setting during the fitting. If the model errors are small, it might even suffice to use a single Jacobian $R_{\Theta}$ for the entire fitting procedure; that is, the same Jacobian can be reused during each iteration.

The analytical Jacobian is computed via \refe{eq:quasi-degeneracy:orbit-response-derivative} from Twiss data which is obtained from the accelerator model evaluated at the current parameter guess. Due to sign convention for quadrupoles, for the vertical dimension the Jacobian needs to be multiplied by \num{-1}.

Using the analytical Jacobian from Twiss data is more efficient than computing the numerical Jacobian since Twiss data is computed only once for the full Jacobian while the numerical approach computes one ORM per quadrupole (i.e. one closed orbit per steerer per quadrupole).
For the BPM and steerer gain parts of the ORM, the analytical equation for the orbit response \refe{eq:orbit-response} is similarly used with Twiss data.

\begin{figure}[hbt]
   \centering
   \includegraphics[width=0.99\textwidth]{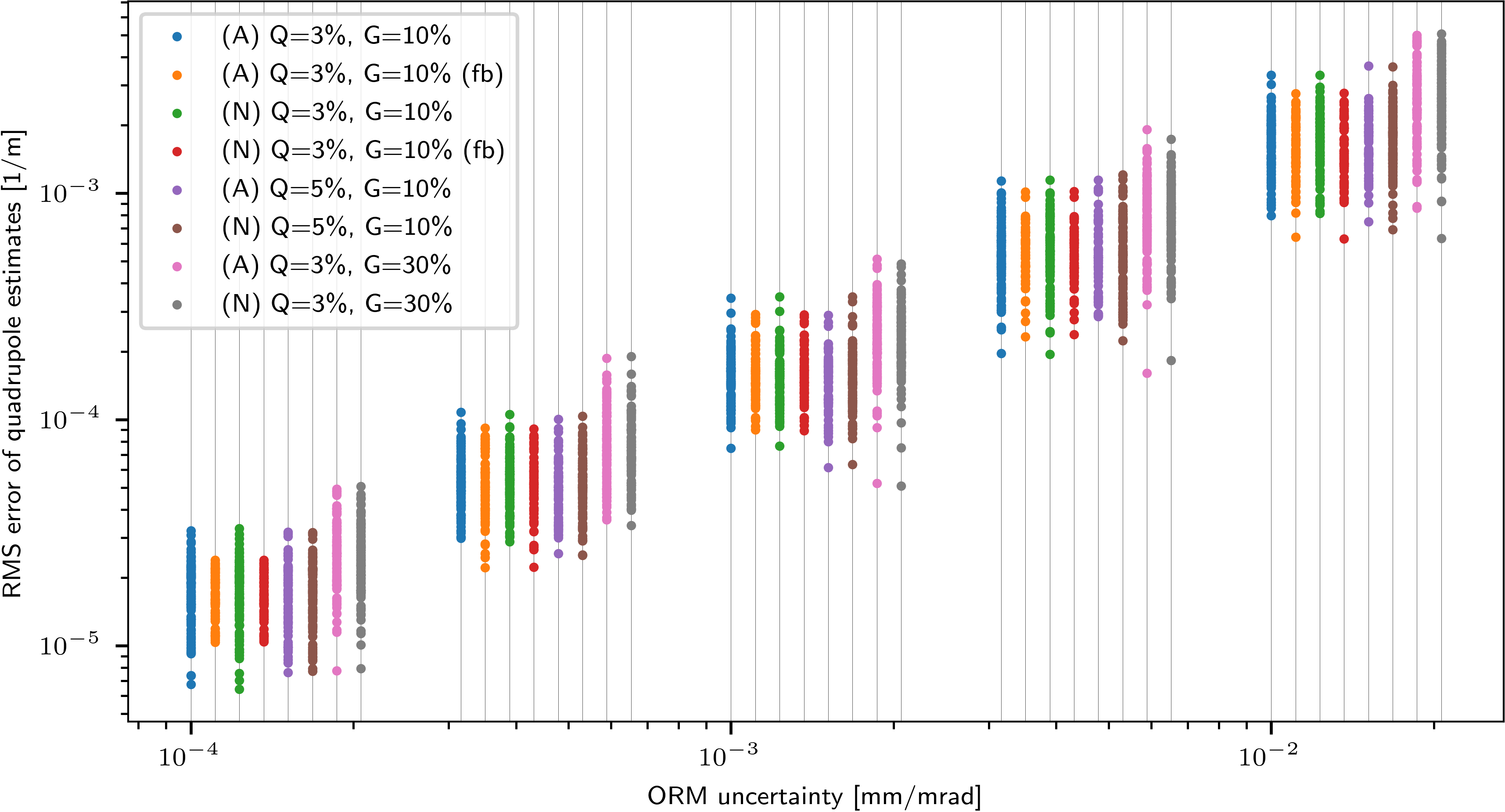}
   \caption{Comparison of simulation results for various cases. \texttt{(A)} and \texttt{(N)} denote, respectively, the usage of analytical or numerical Jacobian. \texttt{Q} and \texttt{G} denote, respectively, the percentage level of random quadrupole and gain errors, uniformly sampled within these bounds. All simulations used Levenberg-Marquardt optimizer except the ones with suffix \texttt{(fb)} which used a purely feedback-like approach using only the analytical or numerical Jacobian obtained for the error-free model optics setting. The feedback-like approach converged for \SI{67}{\percent} of the simulated \texttt{Q=3\%, G=10\%} instances for both, the analytical and numerical Jacobian method. For larger quadrupole or gain errors the rate of convergence decreases further and, hence, these results are not reported. However, simulating quadrupole errors below \SI{2}{\percent} (not shown) results in more than \SI{98}{\percent} convergence rate for the feedback-like approach. The convergence rate does not depend on the simulated ORM uncertainty. All other approaches converge reliably also for the larger error levels shown in the plot. The simulations have been performed for five different ORM uncertainties which are plotted on the horizontal axis: \SIlist{0.1;0.32;1.0;3.2;10.0}{\micro\meter\per\milli\radian}. For each uncertainty level, the eight different cases are shifted horizontally for better visibility (their order from left to right matches the order in the legend from top to bottom); however, each case used the same ORM uncertainty for simulations (the leftmost one). Each uncertainty level contains \num{100} random simulations per case.}
   \label{fig:fitting:simulation-results-comparison}
\end{figure}

Various tests with simulation data have been performed. The tests include random quadrupole and gain errors as well as different levels of simulated orbit response uncertainty. The Levenberg-Marquardt algorithm has been used for the fitting. The results are shown in \reff{fig:fitting:simulation-results-comparison}. It can be observed that the results obtained with the analytical Jacobian match closely with those obtained with the numerical Jacobian. For the simulation case which limits quadrupole errors by \SI{3}{\percent} and gain errors by \SI{10}{\percent}, the feedback-like approach using only the analytical Jacobian obtained for the nominal optics converges in \SI{67}{\percent} of the instances and it reaches unstable lattice configurations for the remaining instances. This is due to the discrepancy of the real Jacobian with respect to the employed Jacobian obtained from nominal optics being too large to allow convergence according to \refe{eq:fitting:feedback-matrix-mult-sequence}. The convergence rate is, however, independent of the simulated ORM uncertainty. For simulated quadrupole errors below \SI{2}{\percent}, the feedback-like approach converges in more than \SI{98}{\percent} of instances. Thus, this approach can be used to correct a lattice that exhibits only small quadrupole drifts over time.
When computing the analytical Jacobian at every iteration of the fitting procedure, it converges and yields good results also for larger simulated model errors as shown in \reff{fig:fitting:simulation-results-comparison}.

\section{Experiment}
\label{sec:experimental-data}

The following experimental data has been collected to support the findings.
ORM and tune measurements have been conducted for two different optics at SIS18: nominal extraction optics and a modified version of the optics by adjusting one of the F-quadrupole families (\texttt{GS01QS1F} family) by $\Delta K_1L = \SI{-1.2d-3}{\per\meter}$ (this quadrupole family includes the F-quadrupoles from the odd numbered sections). Due to very limited experimental time available, beta beating could not be measured, unfortunately. Nevertheless, the tune measurements serve as a verification for the derived quadrupole errors.

The quadrupole errors are estimated with Levenberg-Marquardt optimizer. The analytical Jacobian obtained from Twiss data is used during the optimization. Comparison with results obtained with numerical Jacobian is presented, too. Different approaches for mitigating the quasi-degeneracy are compared as well.

\subsection{Measured data}

\begin{figure}[hbt]
   \centering
   \includegraphics[width=0.49\textwidth]{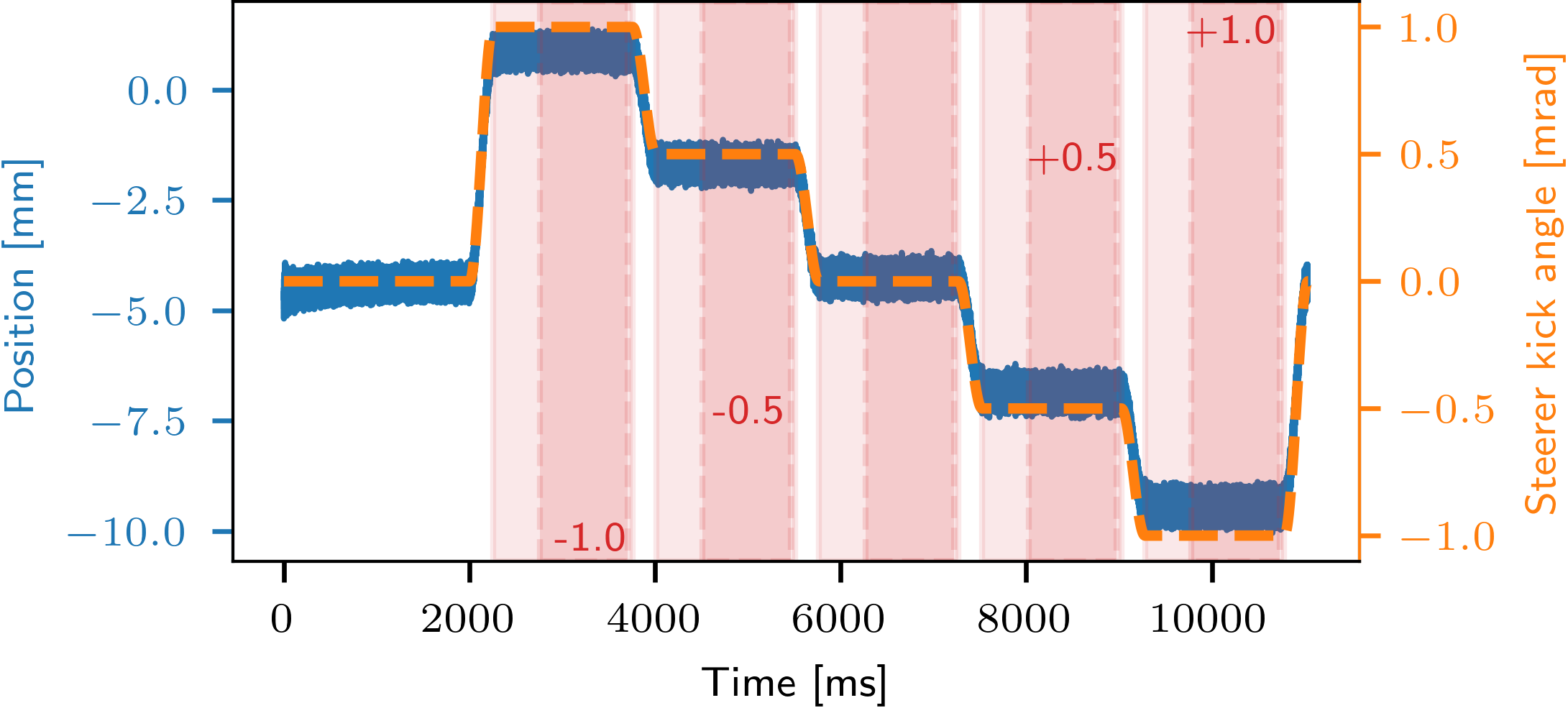}
   \caption{Position data from the horizontal BPM in section \num{1} during measurement of the horizontal steerer in section \num{5}. The steerer setting is overlaid as the dashed curve (the curve is inverted for better visibility). The two vertical axes are not aligned, i.e. there is no meaning in the vertical position of the steerer relative to the position data. The red shaded areas indicate the time windows available for orbit computation. The light shaded area (\SI{500}{\milli\second}) has been excluded because the orbit was still slightly drifting during those time windows. The solid shaded area (\SI{950}{\milli\second}) is used for orbit computation. The white area between two shaded areas is the allocated transition time for the steerer magnets which is \SI{256}{\milli\second}. An additional \SI{2}{\second} are skipped at the beginning of flattop because the horizontal orbit was still drifting during that time window.}
   \label{fig:experiment:position-data-overlaid-by-steerer-values}
\end{figure}

The ORM measurements were performed with \num{5} settings per steerer, \SIlist{-1.0;-0.5;0.0;0.5;1.0}{\milli\radian}, during a long flattop of \SI{11}{\second}. Position data from one of the horizontal BPMs is shown in \reff{fig:experiment:position-data-overlaid-by-steerer-values}. The first \SI{2}{\second} are skipped because the horizontal orbit still drifted during that time window; this is likely because of the bending magnets taking long time to attain their nominal strength. The long flattop duration allowed for long data integration windows of \SI{950}{\milli\second} for each steerer setting in order to reduce the measurement uncertainty. Also, sufficient time, \SI{256}{\milli\second}, was allocated for transitioning between two steerer settings plus an additional \SI{500}{\milli\second} to allow the steerers to attain the new values.
For each machine cycle, the response $r_c$ is computed from a least squares fit of the \num{5} corresponding steerer settings. The final response $r$ is computed as the average over \num{5} subsequent cycles, each inversely weighted with its squared standard error $\sigma_c$ from the least squares fit of the respective response $r_c$:
\begin{equation}
    r = \frac{1}{\sum_{c}\frac{1}{\sigma_c^2}}\sum_{c}\frac{r_c}{\sigma_c^2}
\end{equation}
A measurement uncertainty of \SI{5}{\micro\meter\per\milli\radian} has been reached for the orbit response, with minor variations among the different BPMs. For the measurement of modified optics, the horizontal BPM in section \num{8} malfunctioned and thus had to be removed from the analysis.

\begin{figure}[hbt]
   \centering
   \includegraphics[width=0.51\textwidth]{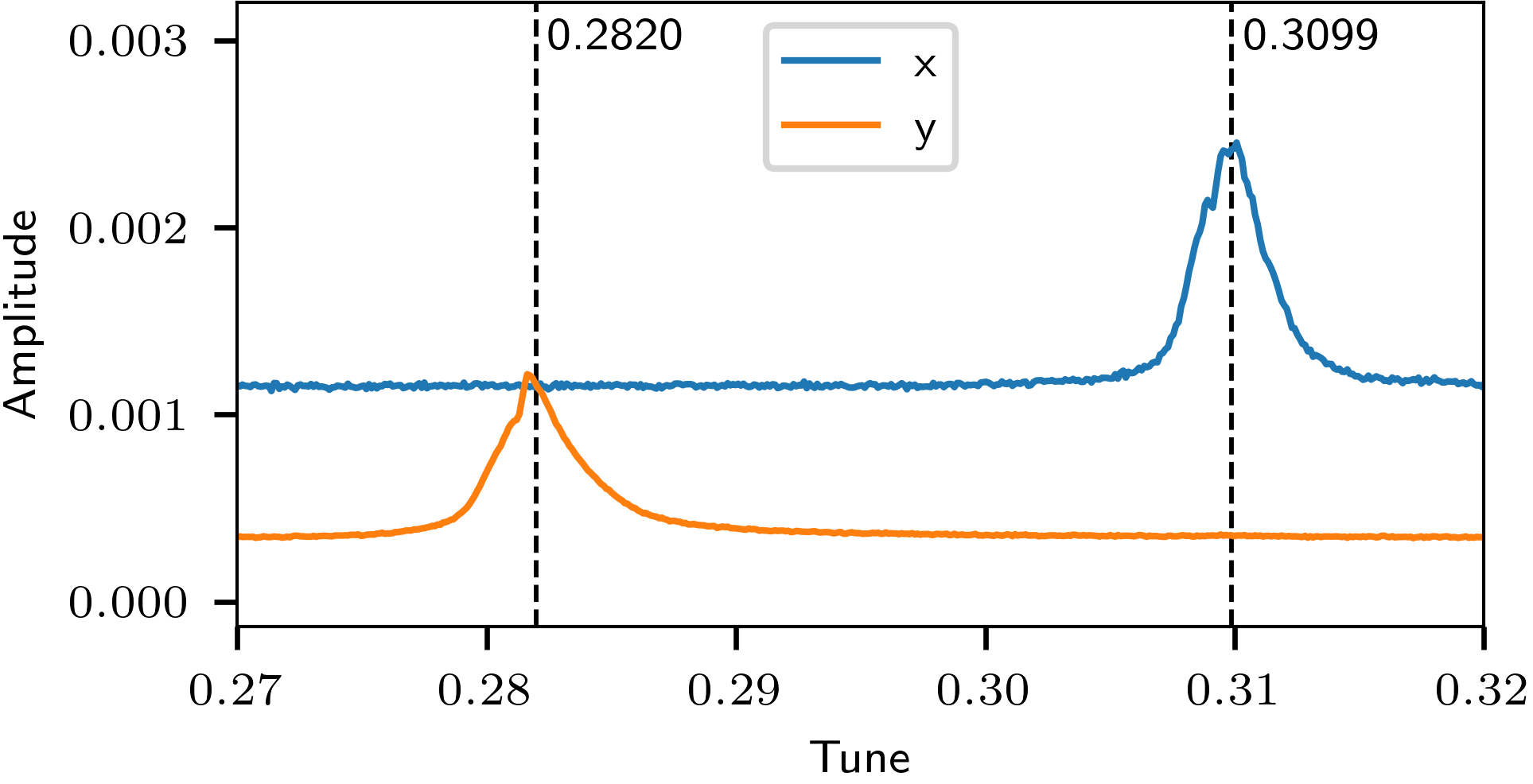}
   \includegraphics[width=0.51\textwidth]{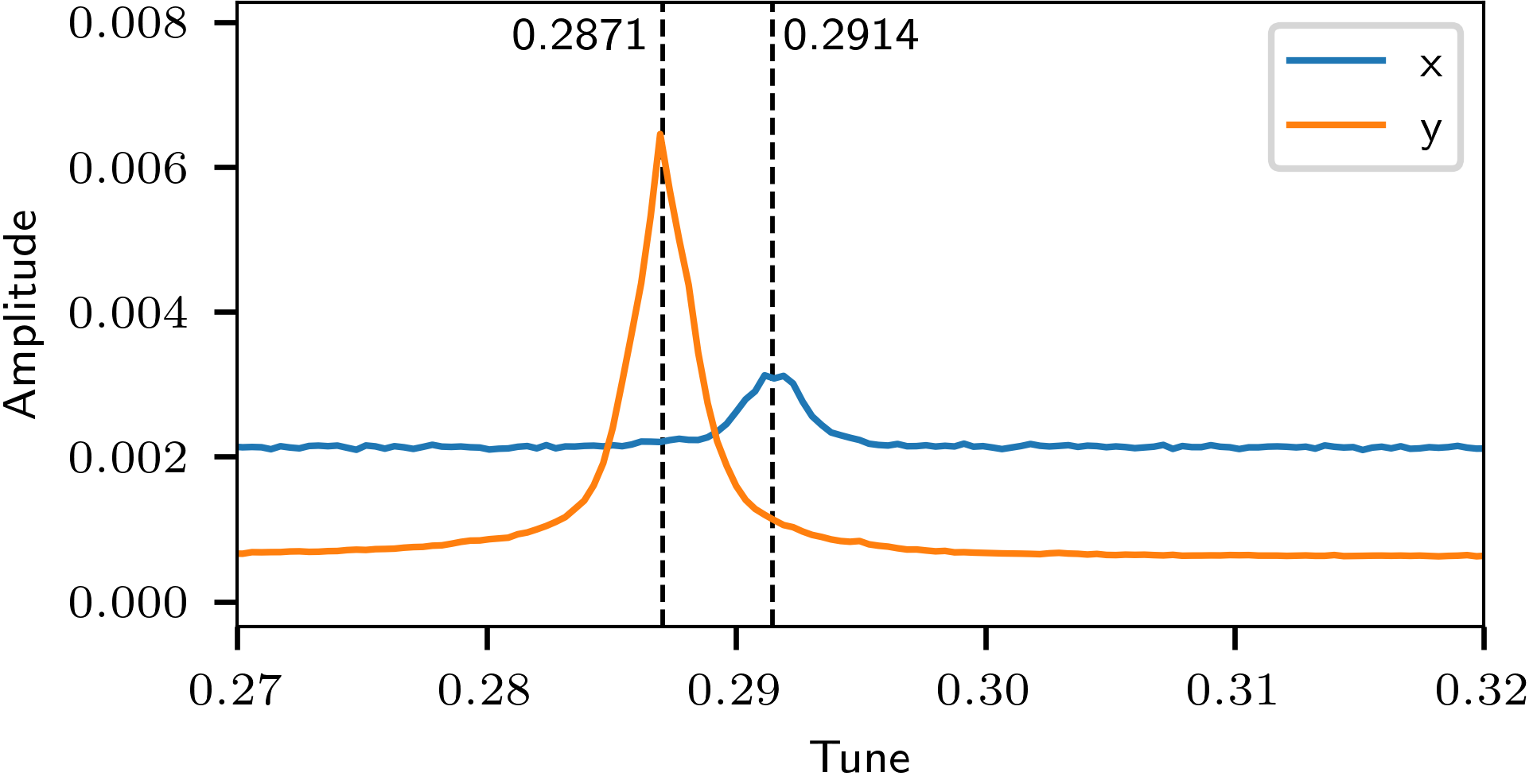}
   \caption{Measured tunes for nominal extraction optics (top) and modified optics (bottom). The model tunes for nominal extraction optics are $0.29$ in both dimensions. The measurement for modified optics was performed on a reduced time scale of \SI{6}{\second} to limit the amount of position data generated.}
   \label{fig:experiment:tune-amplitude-plots}
\end{figure}

Tune measurements have been obtained by excitation via turn by turn stripline exciter and position monitoring. The following values have been measured. The measured tunes are shown in \reff{fig:experiment:tune-amplitude-plots}.

\begin{enumerate}
    \item Nominal extraction optics: \begin{itemize}
        \item $q_h = 0.3099 \pm 0.0014$
        \item $q_v = 0.2820 \pm 0.0011$
    \end{itemize}
    \item Modified extraction optics: \begin{itemize}
        \item $q_h = 0.2914 \pm 0.0008$
        \item $q_v = 0.2871 \pm 0.0007$
    \end{itemize}
\end{enumerate}

\subsection{Mitigation of quasi-degeneracy}

To obtain meaningful results that can be compared, it is important to mitigate the quasi-degeneracy which is mainly driven by the D-T-quadrupole pairs. We compare methods SVD cutoff, adding $\Delta K_1L$ constraints to the Jacobian as well as leaving out T-quadrupoles from the fitting. The removal of T-quadrupoles is justified since they attain small strengths during extraction optics and, thus, much smaller errors are expected for this quadrupole family.
For comparison, we refer to the results obtained without any method for counteracting quasi-degeneracy as \textit{baseline method}.

For each of the methods, we present the difference in estimates between the two optics for the F-quadrupoles; that is, the estimates obtained for modified optics subtracted by the estimates obtained for nominal optics. Both estimates are obtained by starting the fitting procedure from the nominal optics model. Ideally, this difference of estimates should be a zigzag pattern between \SI{-1.2d-3}{\per\meter} and \SI{0}{\per\meter} since the \texttt{GS01QS1F} family contains every second F-quadrupole (i.e. the ones from odd section numbers).

\subsubsection{SVD cutoff}

This is performed as a two stage process. The first stage uses Levenberg-Marquardt to find a (quasi-degenerate) solution for all the involved parameters: quadrupole errors and gain errors. The second stage freezes the thus found gain errors and restarts fitting of quadrupole errors. During each update step, the system $J^T J \Delta p = J^T r$ (where $\Delta p$ is the parameter update and $r$ the residual vector) is solved by computing $(J^T J)^{-1}$ via SVD and truncating a predefined number of smallest singular values to zero. If the SVD spectrum shows a clear drop in the magnitude of singular values then cutting the small singular values will be very efficient. However, for a more flat spectrum, the number of singular values to cut is not obvious and also the resulting estimate might suffer from the truncation. This strongly depends on the use case and the investigated lattice. The optimal cutoff value can be found from simulations, where random orbit uncertainties are cast on the nominal ORM and then inverse modeling with different cutoff values is performed. The one that yields the smallest error in terms of the quadrupole error estimates is then chosen. For our use case, we found that the best results are obtained when the number of cut values is set to \num{11}.

\subsubsection{\texorpdfstring{$\Delta K_1L$ weights}{\textDelta KL weights}}

This approach adds weights to the Jacobian as described in \cite{LOCO:2009:ImprovedFitting}. The purpose of the weights is to limit the amount of change in the $\Delta K_1L$ parameters during each iteration of the fitting process. We determined the pattern of weights $w$ at every iteration by
\begin{equation}
\bm{w} = \sum_{i=1}^{N} \frac{1}{\lambda_i} \bm{v}_i
\end{equation}

where $\lambda_i$ and $\bm{v}_i$ are, respectively, the $i$-th eigenvalue and eigenvector of the $\hat{J}^T \hat{J}$ matrix originating from the Jacobian $\hat{J}$ that represents only the $\Delta K_1L$ parameters and which is evaluated at zero gain errors. Then $w_k$ is the weight for the $k$-th quadrupole. The magnitude of $\bm{w}$ is chosen a priori by a scan over different possible values and then fixed for every iteration. It should be emphasized that for this approach we used the nominal gain Jacobian $\hat{J}$ not only for the computation of the weights but it also replaced the $\Delta K_1L$ part of the actual Jacobian $J$ which is evaluated at the current gain error estimate during each iteration. This is done because when using $J$, the estimated gain errors would obfuscate the degeneracy pattern of the quadrupoles at every iteration. Using $\hat{J}$, on the other hand, allows to directly access the quasi-degeneracy patterns and thus limit them by adding corresponding weights. Using $\hat{J}$ in place of $J$ does not hinder convergence as their agreement is sufficiently close.

\subsubsection{Leaving out T-quadrupoles}

Since the magnitude of T-quadrupole strengths is one order of magnitude smaller than the one of other quadrupoles, their errors are expected to be similarly smaller. Hence, leaving out T-quadrupoles from the fitting will alter the estimates of other quadrupoles (mainly D-quadrupoles) only by a relatively small amount.

\subsubsection{Comparison}

\begin{figure}[hbt]
   \centering
   \includegraphics[width=0.51\textwidth]{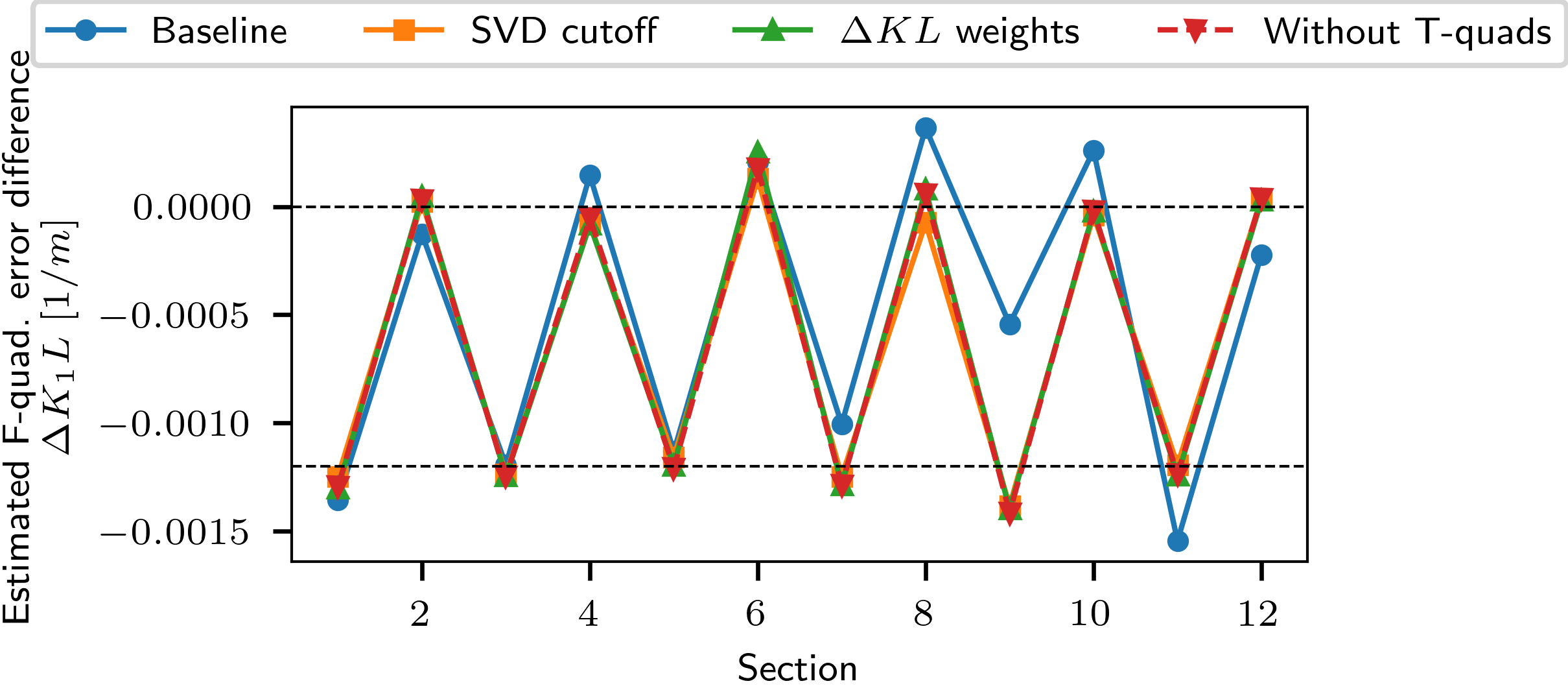}
   \includegraphics[width=0.51\textwidth]{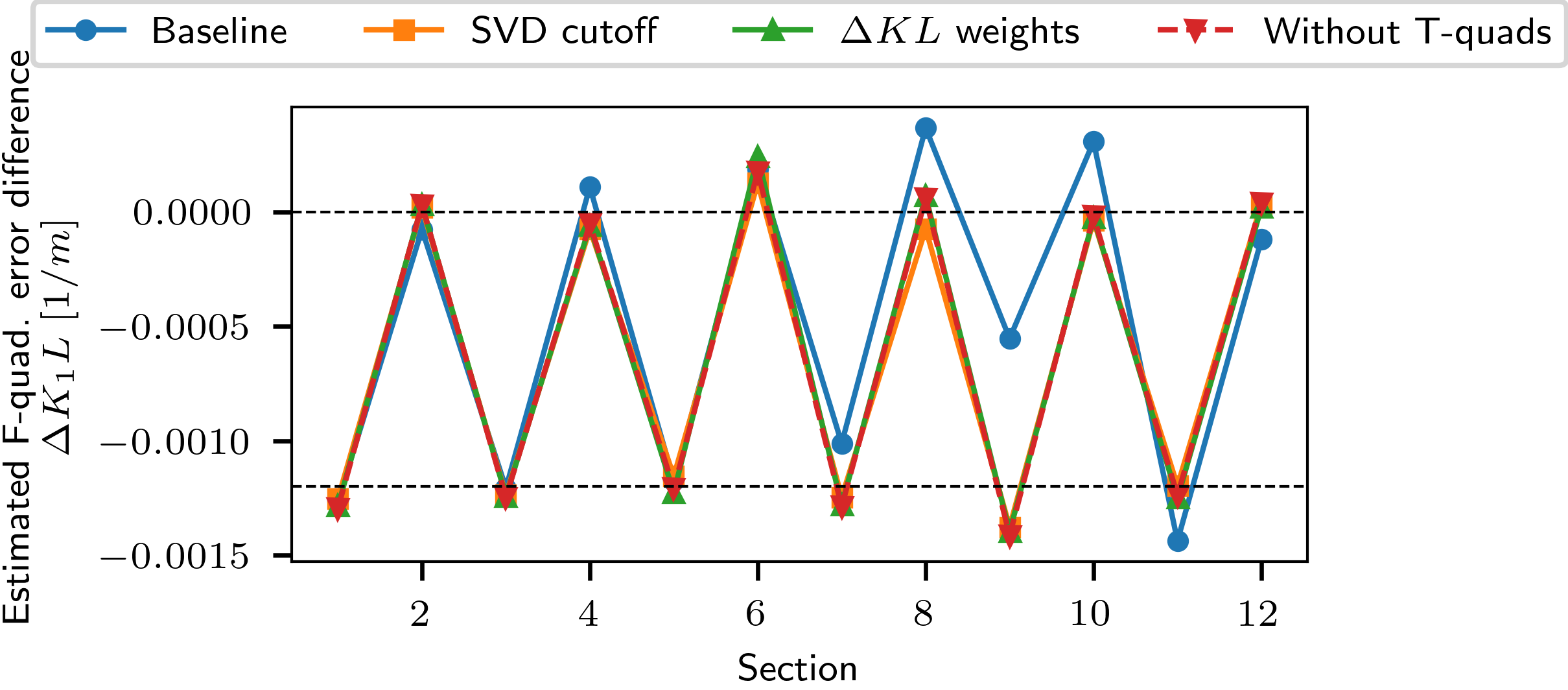}
   \caption{Comparison of inverse modeling results for the F-quadrupoles when using different methods for counteracting quasi-degeneracy. Top: using analytical Jacobian; bottom: using numerical Jacobian. The plots show the difference in estimates for the modified optics and the nominal optics. The two optics differ by the manual adjustment of odd section number F-quadrupoles by $\Delta K_1L = \SI{-1.2d-3}{\per\meter}$. All other quadrupoles, including the F-quadrupoles from even section numbers, have not been modified. The dashed lines indicate the expected (ideal) estimates for the quadrupole errors. The label \textit{Baseline} refers to the results obtained from Levenberg-Marquardt fitting without any countermeasure against the quasi-degeneracy. The error bars due to ORM uncertainty are on the order of \SI{1d-5}{\per\meter} and thus not visible in the plot.}
   \label{fig:experiment:results-comparison-analytical-jacobian}
\end{figure}

Figure~\ref{fig:experiment:results-comparison-analytical-jacobian} shows a comparison between the three abovementioned strategies for counteracting quasi-degeneracy. Since the quasi-degeneracy is mainly driven by the D-T-quadrupole pairs, and T-quadrupoles have a one order of magnitude smaller nominal strength, leaving out the T-quadrupoles from the fit is expected to effectively eliminate the quasi-degeneracy while yielding accurate results (i.e. close to the actual errors). The method of adding $\Delta K_1L$ constraints to the cost function proves similarly efficient as it yields very similar results. The SVD cutoff method shows a slight deviation, mainly because the singular value spectrum is rather flat and removing too many singular values also removes too much information from the Jacobian. The same figure also shows the results obtained with the numerically computed Jacobian. It can be seen that these results closely match the results obtained with the analytical Jacobian. The SVD cutoff method shows a slight deviation between the two methods because the singular value spectrum of the two Jacobian versions is slightly different.

\begin{figure}[hbt]
   \centering
   \includegraphics[width=0.49\textwidth]{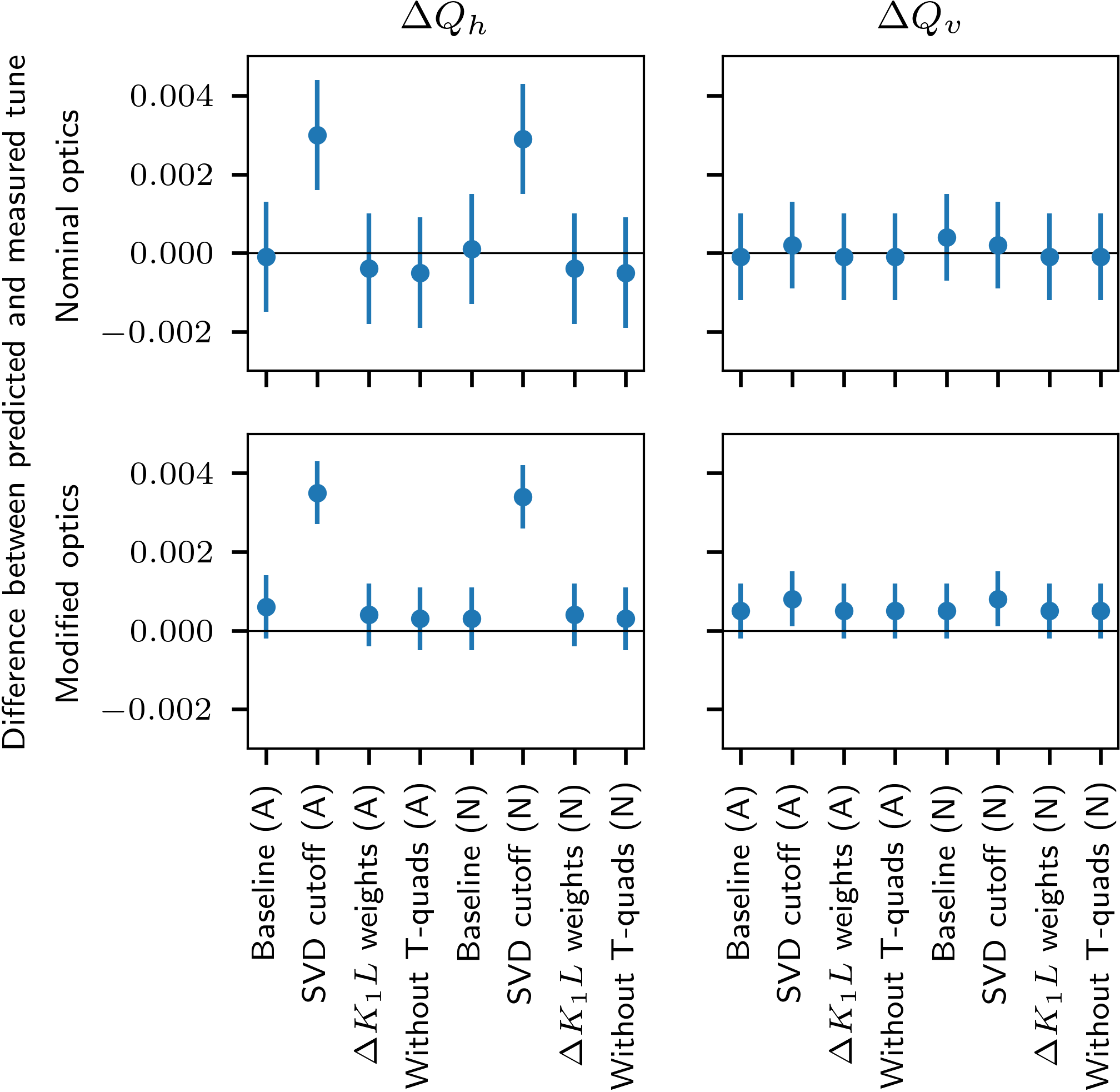}
   \caption{Difference between predicted and measured tunes for the various optics and inverse modeling methods. Top: $\Delta Q_h, \Delta Q_v$ for nominal extraction optics. Bottom: $\Delta Q_h, \Delta Q_v$ for the modified optics. The different methods are indicated on the horizontal axis and are the same for each subplot. \texttt{(A)} and \texttt{(N)} denote, respectively, the usage of analytical and numerical Jacobian. The vertical bars indicate the measurement uncertainty.}
   \label{fig:experiment:tune-plot}
\end{figure}

Table~\ref{tab:experiment:tunes} and \reff{fig:experiment:tune-plot} show an overview of the measured tunes as well as the tunes obtained from the inverse modeling results with the different methods. It can be observed that for all methods except SVD cutoff, the predicted model tunes after fitting match the measured tunes within the measurement uncertainty. The predicted horizontal tune from the SVD cutoff method has a deviation of up to $\approx 3\sigma$ from the measured horizontal tune. This is due to the rather flat singular value spectrum. The agreement of predicted with measured tunes confirms that the fitted models capture the global optics of the real machine. It also emphasizes the effect of quasi-degeneracy, since also the baseline method reproduces the measured tunes closely albeit the $\Delta K_1L$ predictions deviate significantly as can be seen from \reff{fig:experiment:results-comparison-analytical-jacobian}.

\begin{table}[hbt]
\centering
\caption{Resulting tunes from the various fitting methods compared to measured tunes.} 
\begin{tabular}{cc|cccc}
\hline
 & & \multicolumn{2}{c}{Nominal optics} & \multicolumn{2}{c}{Modified optics} \\
 & & $q_h$ & $q_v$ & $q_h$ & $q_v$ \\
\hline
\multirow{2}{*}{Measured} & value & $0.3099$ & $0.2820$ & $0.2914$ & $0.2871$ \\
 & uncertainty & $0.0014$ & $0.0011$ & $0.0008$ & $0.0007$ \\
\hline
\multirow{4}{*}{Analytical Jacobian} & Baseline & 0.3098 & 0.2819 & 0.2920 & 0.2876 \\
 & SVD cutoff & 0.3129 & 0.2822 & 0.2949 & 0.2879 \\
 & $\Delta K_1L$ weights & 0.3095 & 0.2819 & 0.2918 & 0.2876 \\
 & Without T-quads & 0.3094 & 0.2819 & 0.2917 & 0.2876 \\
\hline
\multirow{4}{*}{Numerical Jacobian} & Baseline & 0.3100 & 0.2824 & 0.2917 & 0.2876 \\
 & SVD cutoff & 0.3128 & 0.2822 & 0.2948 & 0.2879 \\
 & $\Delta K_1L$ weights & 0.3095 & 0.2819 & 0.2918 & 0.2876 \\
 & Without T-quads & 0.3094 & 0.2819 & 0.2917 & 0.2876 \\
\hline
\end{tabular}
\label{tab:experiment:tunes}
\end{table}

\section{Conclusions}

The problem of quasi-degeneracy for inverse modeling with orbit response measurements has been studied with regard to the placement of BPMs and steerer magnets, showing that different BPM and steerer placements can noticeably affect the amount of quasi-degeneracy that is present and thus influence the quality of model parameter estimates. These findings emphasize the importance to study the effect of BPM and steerer placements during the design phase of new accelerators.
It has been further shown which BPM and steerer placements cause the inverse problem to be ill-defined which outlines the theoretical limitations of the method. In one dimension, quadrupole triplets surrounded by BPM/steerer, and in both dimensions, quadrupole quintets surrounded by BPMs/steerers cause a rank deficiency in the Jacobian and thus do not have a unique solution in terms of the quadrupole errors.
An analytical version of the Jacobian, relating quadrupole along with BPM and steerer gain errors to the orbit response matrix, has been derived. We have shown that this analytical version, which can be obtained from the lattice's Twiss data, can be used during the fitting in place of the numerically obtained Jacobian. A single Twiss computation for the currently estimated optics is sufficient to construct the analytical Jacobian. This allows for reduced computation time during the fitting procedure, as compared to computing the numerical Jacobian. The approach has been studied with simulation data and also with dedicated measurements at the SIS18 synchrotron at GSI. The presented results suggest the applicability of the method. It should be emphasized that the applicability of the method depends on the accuracy of the thin lens approximation that is used for the derivation of the analytical Jacobian. The better this approximation is, the fewer extra iterations it will require in order to reach convergence at the same level when compared with the numerical Jacobian.
The fitting procedure has been paired with different methods for counteracting quasi-degeneracy. A comparison of the results obtained with the analytical Jacobian to those obtained with the numerical Jacobian shows a close agreement between the two versions. While the methods of adding $\Delta K_1L$ weights or fitting without T-quadrupoles reproduced the measured tunes accurately, the method of SVD cutoff resulted in an increased deviation. This is attributed to the rather flat singular value spectrum of the Jacobian.
The observed tune discrepancies at SIS18 have been explained in light of these findings.
The results of this contribution also provide general hints on the adequate number and placement of steerers and BPMs in favor of a tractable and well conditioned inverse modeling problem. We explored the dependency of quasi-degeneracy on the placement of BPMs and steerers and thus provide insight into how these locations can be chosen for newly designed lattices. Especially for machines where the number of BPMs is limited this can be an important aspect.

\appendix

\section{Derivative of orbit response with respect to quadrupole strength}
\label{seq:appendix:orbit-response-derivative}

Starting with the orbit response $r_{bs}$ induced by steerer $s$ and measured by BPM $b$:
\begin{equation}
    r_{bs} = \underbrace{\sqrt{\beta_b\beta_s}}_{A} \underbrace{\frac{1}{2\sin(\pi Q)}}_{B} \underbrace{\cos(\pi Q - |\mu_b - \mu_s|)}_{C}
\end{equation}

The derivative $\frac{d}{d(K_1L)_k}r_{bs} \equiv r_{kbs}'$ is:
\begin{equation}
    r_{kbs}' = A'BC + AB'C + ABC'
\end{equation}

In the following the individual derivatives $A', B', C'$ are derived.

\begin{equation}
    A' = \frac{1}{2\sqrt{\beta_b\beta_s}}\left[\beta_b'\beta_s + \beta_b\beta_s'\right] \approx -\frac{\beta_k}{2}\sqrt{\beta_b\beta_s}\left[\Psi_{ks} + \Psi_{kb}\right] = -A\frac{\beta_k}{2}\left[\Psi_{ks} + \Psi_{kb}\right]
\end{equation}

where we have used the formula for the beta beating~\cite{SYLee}:
\begin{equation}
    \beta_s' \approx -\beta_s\beta_k\underbrace{\frac{\cos(2\pi Q - 2|\mu_k - \mu_s|)}{2\sin(2\pi Q)}}_{\Psi_{ks}}
\end{equation}
and similarly for $\beta_b', \Psi_{kb}$.

\begin{equation}
    B' = -\frac{1}{2}\frac{\cos(\pi Q)}{\sin(\pi Q)^2}\pi Q' \approx -B\frac{\beta_k}{2}\frac{1}{2\tan(\pi Q)}
\end{equation}

where we have used the formula for the tune change induced by a quadrupolar error~\cite{SYLee}:
\begin{equation}
    Q' \approx \frac{\beta_k}{4\pi}
\end{equation}

\begin{equation}
    C' = -C\tan(\pi Q - |\mu_b - \mu_s|)\left[\frac{\beta_k}{4} - \frac{\mu_{\max} - \mu_{\min}}{|\mu_{\max} - \mu_{\min}|}\left(\mu_{\max}' - \mu_{\min}'\right)\right]
\end{equation}

where we have assumed $\cos(\pi Q - |\mu_b - \mu_s|) \neq 0$ (i.e. $r_{bs} \neq 0$) and reordered the terms $\mu_b, \mu_s$ inside $\cos(\pi Q - |\mu_b - \mu_s|)$ such that the argument of the absolute value is positive, i.e. $|\mu_b - \mu_s| = |\max(\mu_b, \mu_s) - \min(\mu_b, \mu_s)|$ and $\mu_{\max} \equiv \max(\mu_b, \mu_s), \mu_{\min} \equiv \min(\mu_b, \mu_s)$. In that case $\frac{\mu_{\max} - \mu_{\min}}{|\mu_{\max} - \mu_{\min}|} = 1$ and we are only left with the derivative $\mu_{\max}' - \mu_{\min}' = (\mu_{\max} - \mu_{\min})'$. To compute this derivative we consider the change in local phase advance $\Delta\mu_i$ induced by a small quadrupolar error $\Delta(K_1L)_k$~\cite{Lecture:Bruening}:
\begin{equation}
\begin{aligned}
\mu_i &=  \mu_{0,i} + \Delta\mu_i \\
\mu_i &= \int_{s_0}^{s_i}\frac{1}{\beta(\tau)}d\tau + \mu_{s=s_0} \\
&= \int_{s_0}^{s_i}\frac{1}{\beta_0(\tau) + \Delta\beta(\tau)}d\tau + \mu_{s=s_0} \\
&= \int_{s_0}^{s_i}\frac{1}{\beta_0(\tau)}\cdot\frac{1}{1 + \frac{\Delta\beta(\tau)}{\beta_0(\tau)}}d\tau + \mu_{s=s_0} \\
&\approx \int_{s_0}^{s_i}\frac{1}{\beta_0(\tau)}d\tau - \int_{s_0}^{s_i}\frac{\Delta\beta(\tau)}{\beta_0(\tau)^2}d\tau + \mu_{s=s_0} \\
\end{aligned}
\end{equation}
where the subscript $0$ indicates the unperturbed optics functions, i.e. without quadrupole error, and we have used the fact that Taylor series are multiplicative. Considering the difference $\mu_{\max} - \mu_{\min}$ we thus obtain:
\begin{equation}
    \mu_{\max} - \mu_{\min} = \int_{s_{\min}}^{s_{\max}}\frac{1}{\beta_0(\tau)}d\tau - \int_{s_{\min}}^{s_{\max}}\frac{\Delta\beta(\tau)}{\beta_0(\tau)^2}d\tau
\end{equation}
where $s_{\min}, s_{\max}$ denote the corresponding longitudinal lattice positions. Since $\mu_{\max} - \mu_{\min} = \mu_{0,\max} + \Delta\mu_{\max} - \mu_{0,\min} -  \Delta\mu_{\min} = (\mu_{0,\max} - \mu_{0,\min}) + \Delta(\mu_{\max} - \mu_{\min})$ we obtain:
\begin{equation}
    \Delta(\mu_{\max} - \mu_{\min}) = -\int_{s_{\min}}^{s_{\max}}\frac{\Delta\beta(\tau)}{\beta_0(\tau)^2}d\tau
\end{equation}
By using the expression for the beta beating this can be rewritten as:
\begin{equation}
    \Delta(\mu_{\max} - \mu_{\min}) = \Delta(K_1L)_k\frac{\beta_{0,k}}{2\sin(2\pi Q_0)}\int_{s_{\min}}^{s_{\max}}\frac{\cos(2\pi Q_0 - 2|\mu_{0,k} - \mu_0(\tau)|)}{\beta_0(\tau)}d\tau
\end{equation}
Approximating the derivative with $(\mu_{\max} - \mu_{\min})' \approx \frac{\Delta(\mu_{\max} - \mu_{\min})}{\Delta(K_1L)_k}$ and using $\frac{d}{d\tau}\mu_0(\tau) = \frac{1}{\beta_0(\tau)}$ with integration by substitution we obtain:
\begin{equation}
    \mu_{\max}' - \mu_{\min}' = \frac{\beta_{0,k}}{2\sin(2\pi Q_0)} \int_{\mu_{0,\min}}^{\mu_{0,\max}}\cos(2\pi Q_0 - 2|\mu_{0,k} - u|)du
\end{equation}
In the following we drop the subscript $0$ for nominal values, as there is no further ambiguity.

Hence, all derivatives $\{A,B,C\}'$ can be written as $-\{A,B,C\}\frac{\beta_k}{2}f_{\{A,B,C\}}$, i.e the derivative $r_{kbs}'$ can be written as a product of $r_{bs}$, the beta function at the respective quadrupole and a sum of the factors $f_{\{A,B,C\}}$:
\begin{equation}
\label{eq:appendix:orbit-response-derivative}
\begin{aligned}
\frac{dr_{bs}}{d(KL)_k} = -r_{bs}\frac{\beta_k}{2} & \left\{ \frac{1}{2\tan(\pi Q)} + \frac{\tan(\pi Q - |\mu_b - \mu_s|)}{2} \right. \\
& + \Psi_{ks} + \Psi_{kb} \\
& - \left.  \frac{\tan(\pi Q - |\mu_b - \mu_s|)}{\sin(2\pi Q)}\int_{\min(\mu_b,\mu_s)}^{\max(\mu_b,\mu_s)}\cos(2\pi Q - 2|\mu_k - u|)du \right\}
\end{aligned}
\end{equation}

The integral in equation \ref{eq:appendix:orbit-response-derivative} can be solved by taking into account the absolute value function that is part of the integrand. Therefore, we need to divide the integration domain in order to resolve it. For any quadrupole $k$, there are three distinct cases:
\begin{enumerate*}[label=(\Alph*)]
    \item $\mu_{\min} < \mu_{\max} < \mu_k$,
    \item $\mu_{\min} < \mu_k < \mu_{\max}$,
    \item $\mu_k < \mu_{\min} < \mu_{\max}$.
\end{enumerate*}
For cases (A) and (C) the argument of the absolute value assumes the same sign on the entire integration domain and, hence, there is no need to split the integration domain. For case (B) it needs to be split in $[\mu_{\min},\mu_k]$ and $[\mu_k,\mu_{\max}]$.

The solutions are:
\begin{equation}
\label{eq:appendix:integral-term-explicit}
\begin{aligned}
& \int_{\mu_{\min}}^{\mu_{\max}}\cos(2\pi Q - 2|\mu_k - u|)du = \\
= & \begin{cases}
    (A) & \sin(\mu_{\max} - \mu_{\min})\cos(2\pi Q - |\mu_k - \mu_{\max}| - |\mu_k - \mu_{\min}|) \\
    (B) & \sin(|\mu_k - \mu_{\min}|)\cos(2\pi Q - |\mu_k - \mu_{\min}|) + \sin(|\mu_k - \mu_{\max}|)\cos(2\pi Q - |\mu_k - \mu_{\max}|) \\
    (C) & \sin(\mu_{\max} - \mu_{\min})\cos(2\pi Q - |\mu_k - \mu_{\max}| - |\mu_k - \mu_{\min}|) \\
\end{cases}
\end{aligned}
\end{equation}
Hence, the result for cases (A) and (C) is similar and a distinction has to be made between the two different cases (A,C) for which both $\mu_{\min},\mu_{\max}$ are either upstream or downstream of the quadrupole and (B) for which $\mu_{\min}$ is upstream and $\mu_{\max}$ is downstream of the quadrupole.

\section{Derivative of orbit response with respect to quadrupole strength for beamlines}
\label{sec:appendix:orbit-response-derivative-beamlines}

For beamlines, or more generally, non-closed lattices, we have the following formula for the orbit response at BPM $b$ induced by steerer $s$~\cite{OrbitResponseEquationBeamline}:
\begin{equation}
r_{bs} = \begin{cases}
    \sqrt{\beta_b\beta_s}\sin(\mu_b - \mu_s) \quad &, \; \mu_b > \mu_s \\
    0                                              &, \; \mathrm{otherwise} \\
\end{cases}
\end{equation}

The relation for $\frac{\Delta\beta}{\beta}$ for non-closed lattices to first order is given by~\cite{CERN-ACC-NOTE-2018-0025}:
\begin{equation}
\frac{\Delta\beta_x}{\Delta(K_1L)_k} = -\beta_k\beta_x\sin(2\mu_x - 2\mu_k)
\end{equation}
where the subscript $x$ refers to the point of measurement and $k$ refers to the quadrupole; $\mu_x > \mu_k$ is assumed since only downstream regions are affected.

Taking the derivative of $r_{bs}$ with respect to $\Delta(K_1L)_k$ one obtains the following:
\begin{equation}
r_{kbs} \equiv \frac{d r_{bs}}{d \Delta(K_1L)_k} = \begin{cases}
0                                                                                      &, \; \mu_k < \mu_s \\
-r_{bs}\beta_k\frac{\sin(\mu_b - \mu_k)\sin(\mu_k - \mu_s)}{\sin(\mu_b - \mu_s)} \quad &, \; \mu_k > \mu_s \\
\end{cases}
\end{equation}

This can be expanded into $\cos(\mu_k)^2$, $\sin(\mu_k)^2$ and $\cos(\mu_k)\sin(\mu_k)$ terms with their respective coefficient vectors.

Compared with the Jacobian for a circular lattice, the beamline Jacobian additionally has some of its elements zeroed. Thus, the rank of the beamline Jacobian for a given BPM/steerer placement must be less than or equal to the rank of the corresponding circular lattice Jacobian. Our simulations show that it is rank deficient for the cases \texttt{Sh,Sv,Q5+,Bh,Bv} but has full rank for \texttt{Sh,Sv,Q4,Bh,Bv}.

\section{Proof: \texttt{S,Q3,B} Jacobian is rank deficient}
\label{sec:appendix:proof:sqqqb}

The trigonometric expressions in the Jacobian eq.~\ref{eq:quasi-degeneracy:orbit-response-derivative} can be expanded in terms of $\mu_k$ by using the identities $\cos(x\pm y) = \cos(x)\cos(y)\mp\sin(x)\sin(y)$, $\sin(x\pm y) = \sin(x)\cos(y)\pm\cos(x)\sin(y)$, $\sin(2x) = 2\sin(x)\cos(x)$, $\cos(2x) = \cos(x)^2 - \sin(x)^2$, $1 = \cos(x)^2 + \sin(x)^2$. The resulting expression can be grouped by terms containing $\cos(\mu_k)^2$, $-\sin(\mu_k)^2$ and $2\cos(\mu_k)\sin(\mu_k)$. This allows to represent each column of the Jacobian by a set of three coefficient vectors, one for each of the trigonometric terms. These coefficient vectors contain the phase advances of BPMs/steerers and their structure only depends on whether the BPM/steerer placement is of type A ($\mu_{\min} < \mu_{\max} < \mu_k$), type B ($\mu_{\min} < \mu_k < \mu_{\max}$) or type C ($\mu_k < \mu_{\min} < \mu_{\max}$), where $\mu_{\min}\equiv\min(\mu_b,\mu_s)$ and $\mu_{\max}\equiv\max(\mu_b,\mu_s)$. Since the quadrupole triplets of \texttt{S,Q3,B} are not interleaved by BPMs/steerers, the structure of coefficient vectors is the same for each quadrupole in a triplet. In fact, these three coefficient vectors can be used for more than three consecutive quadrupoles as well since the coefficient vectors only need to be multiplied by the three trigonometric factors containing $\mu_k$ for a given quadrupole in order to generate the corresponding column of the Jacobian. Hence, this proof applies to \texttt{S,Q3+,B} BPM/steerer placements as well. Thus, one set of three coefficient vectors is sufficient to generate the Jacobian columns for a full quadrupole $n$-tuplet with $n\geq 3$. This means that there are a total of $3N$ coefficient vectors, one $3$-tuple per quadrupole $n$-tuplet in each of the $N$ sections. These column vectors form the column span of any \texttt{S,Qn+,B} Jacobian for $n \geq 3$. The structure of these coefficient vectors, in terms of the phase advance types A, B, C, is shown exemplary for $N=4, n=3$ in schematic \ref{prog:jacobian-columns-types-abc}.

\begin{figure}
\begin{verbatim}
section:   1   2   3   4 
   quad:  FDT FDT FDT FDT
------------- --- --- ---
[1]<1>    BBB AAA AAA AAA 
[1]<2>    CCC AAA AAA AAA 
[1]<3>    CCC BBB AAA AAA 
[1]<4>    CCC BBB BBB AAA 
[2]<1>    BBB BBB AAA AAA 
[2]<2>    CCC BBB AAA AAA 
[2]<3>    CCC CCC AAA AAA 
[2]<4>    CCC CCC BBB AAA 
[3]<1>    BBB BBB BBB AAA 
[3]<2>    CCC BBB BBB AAA 
[3]<3>    CCC CCC BBB AAA 
[3]<4>    CCC CCC CCC AAA 
[4]<1>    BBB BBB BBB BBB 
[4]<2>    CCC BBB BBB BBB 
[4]<3>    CCC CCC BBB BBB 
[4]<4>    CCC CCC CCC BBB
\end{verbatim}
\caption{This schematic shows the Jacobian elements' types A,B,C for $N=4$ sections and $n=3$ quadrupoles forming a triplet in each of the sections. The quadrupoles in a triplet are labeled \texttt{F,D,T}. \texttt{[i]} stands for the $i$th BPM and \texttt{<i>} stands for the the $i$th steerer. As can be seen, the quadrupoles within a triplet all share the same type for each BPM/steerer pair.}
\label{prog:jacobian-columns-types-abc}
\end{figure}

We use the following set of abbreviations to simplify the notation:
\begin{equation}
\begin{aligned}
u &\equiv \mu_{\max} + \mu_{\min} \\
v &\equiv \mu_{\max} - \mu_{\min} \\
T &\equiv \tan(\pi Q - |\mu_{\max} - \mu_{\min}|) = \tan(\pi Q - v) \\
\tilde{T} &\equiv \frac{1}{2\tan(\pi Q)} + \frac{T}{2} \\
\end{aligned}
\end{equation}

Further, (1) is used to represent $\cos(\mu_k)^2$, (2) for $-\sin(\mu_k)^2$ and (3) for $2\cos(\mu_k)\sin(\mu_k)$.

The specific expressions for the coefficient vectors, in dependence on the trigonometric factor (1,2,3) and type (A,B,C), are shown in table~\ref{tab:appendix:proof:coefficient-vector-expressions}.

\begin{table}[hbt]
\centering
\caption{Expressions for the coefficient vectors for the different types A,B,C. The relationship $\cos(x) + \cos(y) = 2\cos(\frac{x+y}{2})\cos(\frac{x-y}{2})$ has been used to combine the $\cos$ terms originating from the $\Psi_{ks}$ and $\Psi_{kb}$ terms. Note that for each (A,B,C), the only difference in the $(1)$ and $(2)$ expressions is the sign of the trailing terms.}
\begin{tabular}{|c|c|c|}
\hline
\multirow{3}{*}{$(1)$} & A & $2\cos(2\pi Q + u)\left[\cos(v) - T\sin(v)\right] + \tilde{T}$ \\
\cline{2-3}
 & B & $2\cos(u)\left[\cos(2\pi Q - v) + T\sin(2\pi Q - v)\right] - 2T\sin(2\pi Q) + \tilde{T}$ \\
\cline{2-3}
 & C & $2\cos(2\pi Q - u)\left[\cos(v) - T\sin(v)\right] + \tilde{T}$ \\
\hline
\multirow{3}{*}{$(2)$} & A & $2\cos(2\pi Q + u)\left[\cos(v) - T\sin(v)\right] - \tilde{T}$ \\
\cline{2-3}
 & B & $2\cos(u)\left[\cos(2\pi Q - v) + T\sin(2\pi Q - v)\right] + 2T\sin(2\pi Q) - \tilde{T}$ \\
\cline{2-3}
 & C & $2\cos(2\pi Q - u)\left[\cos(v) - T\sin(v)\right] - \tilde{T}$ \\
\hline
\multirow{3}{*}{$(3)$} & A & $2\sin(2\pi Q + u)\left[\cos(v) - T\sin(v)\right]$ \\
\cline{2-3}
 & B & $2\sin(u)\left[\cos(2\pi Q - v) + T\sin(2\pi Q - v)\right]$ \\
\cline{2-3}
 & C & $-2\sin(2\pi Q - u)\left[\cos(v) - T\sin(v)\right]$ \\
\hline
\end{tabular}
\label{tab:appendix:proof:coefficient-vector-expressions}
\end{table}

The expressions in table~\ref{tab:appendix:proof:coefficient-vector-expressions} can be further simplified by noting the following relationships:
\begin{equation}
\begin{aligned}
\cos(v) - T\sin(v) &= \frac{\cos(\pi Q)}{\cos(\pi Q - v)} \\
\cos(2\pi Q - v) + T\sin(2\pi Q - v) &= \frac{\cos(\pi Q)}{\cos(\pi Q - v)} \\
\tilde{T} &= \frac{\cos(\pi Q)}{\cos(\pi Q - v)}\frac{\cos(v)}{2\sin(\pi Q)\cos(\pi Q)}
\end{aligned}
\end{equation}

Thus, $2\frac{\cos(\pi Q)}{\cos(\pi Q - v)}$ is a common factor for all expression in table~\ref{tab:appendix:proof:coefficient-vector-expressions} and removing this factor does not alter the rank of the matrix. We therefore obtain the simplified expressions shown in table~\ref{tab:appendix:proof:coefficient-vector-expressions-simplified}.

\begin{table}[hbt]
\centering
\caption{Simplified expressions for the coefficient vectors for the different types (A,B,C). The common factor $2\frac{\cos(\pi Q)}{\cos(\pi Q - v)}$ has been removed from the expressions in table~\ref{tab:appendix:proof:coefficient-vector-expressions}.} 
\begin{tabular}{|c|c|c|}
\hline
\multirow{3}{*}{$(1)$} & A & $\cos(2\pi Q + u) + \frac{\cos(v)}{4\sin(\pi Q)\cos(\pi Q)}$ \\
\cline{2-3}
 & B & $\cos(u) + \left[\frac{\cos(v)}{4\sin(\pi Q)\cos(\pi Q)} - \frac{\sin(2\pi Q)\sin(\pi Q - v)}{\cos(\pi Q)}\right]$ \\
\cline{2-3}
 & C & $\cos(2\pi Q - u) + \frac{\cos(v)}{4\sin(\pi Q)\cos(\pi Q)}$ \\
\hline
\multirow{3}{*}{$(2)$} & A & $\cos(2\pi Q + u) - \frac{\cos(v)}{4\sin(\pi Q)\cos(\pi Q)}$ \\
\cline{2-3}
 & B & $\cos(u) - \left[\frac{\cos(v)}{4\sin(\pi Q)\cos(\pi Q)} - \frac{\sin(2\pi Q)\sin(\pi Q - v)}{\cos(\pi Q)}\right]$ \\
\cline{2-3}
 & C & $\cos(2\pi Q - u) - \frac{\cos(v)}{4\sin(\pi Q)\cos(\pi Q)}$ \\
\hline
\multirow{3}{*}{$(3)$} & A & $\sin(2\pi Q + u)$ \\
\cline{2-3}
 & B & $\sin(u)$ \\
\cline{2-3}
 & C & $-\sin(2\pi Q - u)$ \\
\hline
\end{tabular}
\label{tab:appendix:proof:coefficient-vector-expressions-simplified}
\end{table}

Let $\tilde{J}$ be the column-wise stack of the $3N$ coefficient vectors emerging from the simplified expressions in table~\ref{tab:appendix:proof:coefficient-vector-expressions-simplified}. Since all the used simplifications preserved the column span of the Jacobian (up to constant factors), the nullspace and thus the rank of $\tilde{J}$ is similar to that of the original Jacobian $J$. Thus, it is sufficient to show that $\tilde{J}$ is rank deficient, i.e. that there exists a vector $\vec{v}$ such that $\tilde{J}\cdot\vec{v} = \vec{0}$. This matrix multiplication involves the row-wise summation of the various coefficient vectors that make up the matrix $\tilde{J}$. Each row contains at most the three distinct types A,B,C (see schematic \ref{prog:jacobian-columns-types-abc}). Thus, each row-wise sum is of the form $\sum_{X\in\{A,B,C\}}\rho_X\cdot\{(1),X\} + \sum_{X\in\{A,B,C\}}\sigma_X\cdot\{(2),X\} + \sum_{X\in\{A,B,C\}}\tau_X\cdot\{(3),X\}$ where $\rho_{X}$ stands for the sum of entries in $\vec{v}$ corresponding to type $\{(1),X\}$ in the coefficient matrix and similarly $\sigma$ refers to type (2) and $\tau$ to type (3).
If we require $\sum_{X\in\{A,B,C\}}(\rho_X - \sigma_X) = 0$ then the terms involving $\frac{\cos(v)}{4\sin(\pi Q)\cos(\pi Q)}$ in table~\ref{tab:appendix:proof:coefficient-vector-expressions-simplified} vanish. Thus, we can create a further simplified matrix that consists of the expressions in table~\ref{tab:appendix:proof:coefficient-vector-expressions-simplified} with these terms removed and augmented by an additional row which enforces the condition $\sum_{X\in\{A,B,C\}}(\rho_X - \sigma_X) = 0$ which allowed the removal of those terms. The new version is shown in table~\ref{tab:appendix:proof:coefficient-vector-expressions-simplified-2}. It should be noted that this is not an equivalence transformation, but the nullspace of the new matrix is contained in the nullspace of the original matrix. Hence, it is sufficient to show that the new matrix represented by table~\ref{tab:appendix:proof:coefficient-vector-expressions-simplified-2} is rank deficient.

\begin{table}[hbt]
\centering
\caption{Further simplified expressions for the coefficient vectors for the different types A,B,C. The additional requirement $\sum_{X\in\{A,B,C\}}(\rho_X - \sigma_X) = 0$ has to be satisfied.} 
\begin{tabular}{|c|c|c|}
\hline
\multirow{3}{*}{$(1)$} & A & $\cos(2\pi Q + u)$ \\
\cline{2-3}
 & B & $\cos(u) - \frac{\sin(2\pi Q)}{\cos(\pi Q)}\sin(\pi Q - v)$ \\
\cline{2-3}
 & C & $\cos(2\pi Q - u)$ \\
\hline
\multirow{3}{*}{$(2)$} & A & $\cos(2\pi Q + u)$ \\
\cline{2-3}
 & B & $\cos(u) + \frac{\sin(2\pi Q)}{\cos(\pi Q)}\sin(\pi Q - v)$ \\
\cline{2-3}
 & C & $\cos(2\pi Q - u)$ \\
\hline
\multirow{3}{*}{$(3)$} & A & $\sin(2\pi Q + u)$ \\
\cline{2-3}
 & B & $\sin(u)$ \\
\cline{2-3}
 & C & $-\sin(2\pi Q - u)$ \\
\hline
\end{tabular}
\label{tab:appendix:proof:coefficient-vector-expressions-simplified-2}
\end{table}

We can reorder the various terms of $\tilde{J}$ to construct a new matrix $\tilde{M}$ such that the columns of $\tilde{M}$ correspond to $\rho_i + \sigma_i$, $\tau_i$ and $\rho_i - \sigma_i$ (in that order) where $i$ refers to the $i$-th column of the three matrices containing all type-(1,2,3) terms. This reordering preserves the dot product $\tilde{J}\cdot\vec{v} = \tilde{M}\cdot\vec{v}$. Only the $\rho_i - \sigma_i$ terms depend on $v$ while the other terms depend on $u$. The overall matrix thus consists of a column-wise stack of three sub-matrices corresponding to $\rho_i + \sigma_i$, $\tau_i$ and $\rho_i - \sigma_i$ and has the following form:
\begin{equation}\label{eq:appendix:proof:M-tilde}
\tilde{M} = \begin{bmatrix}
& & & & & & & & \\
& M^{\rho + \sigma} & & & M^{\tau} & & & M^{\rho - \sigma} & \\
& & & & & & & & \\
0 & \dots & 0 & 0 & \dots & 0 & 1 & \dots & 1 \\
\end{bmatrix}
\end{equation}

The additional last row enforces the condition $\sum_{X\in\{A,B,C\}}(\rho_X - \sigma_X) = 0$. While the original Jacobian $J$ has shape $N^2\times 3N$ (for $N$ sections), the new matrix $\tilde{M}$ has shape $(N^2+1)\times 3N$. By the above derivation it has, however, the same nullspace as $J$. Thus, it is sufficient to show that $\tilde{M}$ is rank deficient. Because the rank of a matrix does not change under row- or column-wise multiplication with a nonzero constant, the common factor $\frac{\sin(2\pi Q)}{\cos(\pi Q)}$ can be removed from the $M^{\rho - \sigma}$ matrix leaving it with only $\sin(\pi Q - v)$ terms.

Since the Gram matrix $A^T A$ of any $m\times n$ matrix $A$ ($m\geq n$) has the same rank as the original matrix $A$, it is sufficient to show that the Gram matrix of $\tilde{M}$ is rank deficient. Since the Gram matrix is a square matrix, its determinant can be computed from the original matrix via the Cauchy-Binet formula~\cite{LinearAlgebra:CauchyBinet}:
\begin{equation}\label{eq:appendix:proof:cauchy-binet}
\det(\tilde{M}^T \tilde{M}) = \sum_{\alpha\in\mathrm{INC}(m,n)}\det\left(\tilde{M}\left[\alpha|\ubar{n}\right]\right)^2 = 0
\end{equation}
where $\ubar{n}$ denotes the set of numbers $\{1,2,\dots,n\}$ and $\mathrm{INC}(m,n)$ denotes the set of all strictly increasing functions from $\ubar{m}$ to $\ubar{n}$; $\tilde{M}\left[\alpha|\ubar{n}\right]$ denotes the sub-matrix of $\tilde{M}$ that emerges from selecting the rows with indices given by $\alpha$ and column indices given by $\ubar{n}$.

Equation~\ref{eq:appendix:proof:cauchy-binet} implies that the determinants of all individual sub-matrices $\tilde{M}\left[\alpha|\ubar{n}\right]$ need to be zero.

To further simplify the involved expressions, we make use of the identities $\cos(x) = \frac{1}{2}\left(e^{ix}+e^{-ix}\right)$ and $\sin(x) = \frac{1}{2i}\left(e^{ix}-e^{-ix}\right)$ which allow to replace the various $\cos,\sin$ terms with the following expressions:
\begin{equation}\label{eq:appendix:proof:cos-sin-as-exp}
\begin{aligned}
\cos(\mu_{\max} + \mu_{\min}) &= \frac{p^2q^2 + 1}{2pq} \\
\cos(2\pi Q + \mu_{\max} + \mu_{\min}) &= \frac{p^2q^2g^4 + 1}{2pqg^2} \\
\cos(2\pi Q - \mu_{\max} - \mu_{\min}) &= \frac{p^2q^2 + g^4}{2pqg^2} \\
\sin(\mu_{\max} + \mu_{\min}) &= \frac{p^2q^2 - 1}{2ipq} \\
\sin(2\pi Q + \mu_{\max} + \mu_{\min}) &= \frac{p^2q^2g^4 - 1}{2ipqg^2} \\
\sin(2\pi Q - \mu_{\max} - \mu_{\min}) &= \frac{p^2q^2 - c^4}{2ipqg^2} \\
\sin(\pi Q - \mu_{\max} + \mu_{\min}) &= \frac{q^2g^2 - p^2}{2ipqg} \\
\end{aligned}
\end{equation}
where $p \equiv e^{i\mu_{\max}}$, $q \equiv e^{i\mu_{\min}}$, $g \equiv e^{i\pi Q}$ for the given values of $\mu_{\max}, \mu_{\min}$ in each row.

It is sufficient to show the rank deficiency for the $N=3, n=3$ (i.e. 3 sections containing quadrupole triplets) case; the general case $N>3$ follows from the symmetric placement of lattice elements from one section to another and $n>3$ follows from the fact that the same set of three coefficient vectors is sufficient to generate the Jacobian columns of any quadrupole $n$-tuplet, i.e. $\tilde{M}$ is a $(N^2+1)\times 3N$ matrix independent of $n$.

The expressions in \refe{eq:appendix:proof:cos-sin-as-exp} can be further simplified by multiplying columns 1,2,3 of $\tilde{M}$ (containing only $\cos$ terms) by $2g^2$, columns 4,5,6 (containing only $\sin$ terms) by $2ig^2$ and columns 7,8,9 (containing only $\sin$ terms) by $2ig$. Then the first row can be multiplied by $(2ig)^{-1}$ and each other row can be multiplied by their respective $pq$ whose inverse occurs in every element across a row. Note that these elementary row/column operations preserve the rank of the matrix. This yields the further simplified expressions given by:
\begin{equation}\label{eq:appendix:proof:cos-sin-as-exp-2}
\begin{aligned}
\cos(\mu_{\max} + \mu_{\min}) &\rightarrow p^2q^2g^2 + g^2 \\
\cos(2\pi Q + \mu_{\max} + \mu_{\min}) &\rightarrow p^2q^2g^4 + 1 \\
\cos(2\pi Q - \mu_{\max} - \mu_{\min}) &\rightarrow p^2q^2 + g^4 \\
\sin(\mu_{\max} + \mu_{\min}) &\rightarrow p^2q^2g^2 - g^2 \\
\sin(2\pi Q + \mu_{\max} + \mu_{\min}) &\rightarrow p^2q^2g^4 - 1 \\
\sin(2\pi Q - \mu_{\max} - \mu_{\min}) &\rightarrow p^2q^2 - g^4 \\
\sin(\pi Q - \mu_{\max} + \mu_{\min}) &\rightarrow q^2g^2 - p^2 \\
\end{aligned}
\end{equation}

Thus, the resulting matrix, with $\cos,\sin$ terms being replaced by \refe{eq:appendix:proof:cos-sin-as-exp-2}, contains only various polynomial terms as elements. With the help of a computer algebra system such as PARI/GP~\cite{PARIGP} it can be shown that the determinants of all $9\times 9$ sub-matrices of the simplified $10\times 9$ matrix $\tilde{M}$ are identical to zero. From this follows that $\tilde{M}$ is rank deficient, according to \refe{eq:appendix:proof:cauchy-binet}. An example program is given by program \ref{prog:appendix:proof:pari-gp}.

\begin{figure}
\begin{Verbatim}[fontsize=\scriptsize]
compute() = 
{
M = [0,0,0,0,0,0,1,1,1;
a^2*d^2*g^2+g^2,a^2*d^2*g^4+1,a^2*d^2*g^4+1,a^2*d^2*g^2-g^2,a^2*d^2*g^4-1,a^2*d^2*g^4-1,a^2*g^2-d^2,0,0;
a^2*e^2*g^2+g^2,a^2*e^2*g^2+g^2,a^2*e^2*g^4+1,a^2*e^2*g^2-g^2,a^2*e^2*g^2-g^2,a^2*e^2*g^4-1,a^2*g^2-e^2,a^2*g^2-e^2,0;
a^2*f^2*g^2+g^2,a^2*f^2*g^2+g^2,a^2*f^2*g^2+g^2,a^2*f^2*g^2-g^2,a^2*f^2*g^2-g^2,a^2*f^2*g^2-g^2,a^2*g^2-f^2,a^2*g^2-f^2,a^2*g^2-f^2;
d^2*b^2+g^4,d^2*b^2*g^4+1,d^2*b^2*g^4+1,d^2*b^2-g^4,d^2*b^2*g^4-1,d^2*b^2*g^4-1,0,0,0;
d^2*c^2+g^4,d^2*c^2*g^2+g^2,d^2*c^2*g^4+1,d^2*c^2-g^4,d^2*c^2*g^2-g^2,d^2*c^2*g^4-1,0,d^2*g^2-c^2,0;
b^2*e^2+g^4,b^2*e^2*g^2+g^2,b^2*e^2*g^4+1,b^2*e^2-g^4,b^2*e^2*g^2-g^2,b^2*e^2*g^4-1,0,b^2*g^2-e^2,0;
b^2*f^2+g^4,b^2*f^2*g^2+g^2,b^2*f^2*g^2+g^2,b^2*f^2-g^4,b^2*f^2*g^2-g^2,b^2*f^2*g^2-g^2,0,b^2*g^2-f^2,b^2*g^2-f^2;
e^2*c^2+g^4,e^2*c^2+g^4,e^2*c^2*g^4+1,e^2*c^2-g^4,e^2*c^2-g^4,e^2*c^2*g^4-1,0,0,0;
c^2*f^2+g^4,c^2*f^2+g^4,c^2*f^2*g^2+g^2,c^2*f^2-g^4,c^2*f^2-g^4,c^2*f^2*g^2-g^2,0,0,c^2*g^2-f^2];
for (row = 1, 10, print(matdet(M[^row,])));
}
\end{Verbatim}
\caption{PARI/GP program for verifying that the determinant of every $9\times 9$ sub-matrix of the $10\times 9$ $\tilde{M}$ matrix for the $N=3$ case is identical to zero. The simplifications from \ref{eq:appendix:proof:cos-sin-as-exp-2} have been applied. The following abbreviations are used: $\{a,b,c\} \equiv e^{i\mu_{b,\{1,2,3\}}}$, $\{d,e,f\} \equiv e^{i\mu_{s,\{1,2,3\}}}$, $g \equiv e^{i\pi Q}$. PARI/GP version 2.13.4 has been used. The program can be run by copying it into a file \texttt{main.gp} and then running \texttt{path/to/gp2c-run main.gp} followed by typing \texttt{compute()}.}
\label{prog:appendix:proof:pari-gp}
\end{figure}

It is worth noting that the proof does not make any assumptions on the values of $\mu_{b,j}, \mu_{s,j}, Q$. Thus, the rank deficiency holds for arbitrary values of $\mu_{b,j}, \mu_{s,j}, Q$ and does not restrict the optics nor the specific placement of BPMs or steerers in terms of their phase advance.

\section{Proof: \texttt{Sh,Sv,Q6,Bh,Bv} Jacobian is rank deficient}
\label{sec:appendix:proof:ssq6bb}

The proof for the \texttt{Sh,Sv,Q6,Bh,Bv} placement is analogous to the one obtained for \texttt{S,Q3,B} (appendix \ref{sec:appendix:proof:sqqqb}). Instead of three coefficient vectors there are six coefficient vectors, three for each dimension. These coefficient vectors are orthogonal since the horizontal coefficient vectors only have nonzero entries in the horizontal part of the Jacobian while the vertical coefficient vectors only have nonzero entries in the vertical part of the Jacobian and the two parts of the Jacobian are entirely separate. Hence, we can construct a matrix similar to $\tilde{M}$ in \refe{eq:appendix:proof:M-tilde} but now the matrix is a block diagonal of shape $(2N^2+2)\times 6N$ where the left upper block is the $\tilde{M}$ for the horizontal dimension and the right lower block is the $\tilde{M}$ for the vertical dimension. Both blocks independently induce a rank deficiency as shown in \ref{sec:appendix:proof:sqqqb}. Thus, the rank deficiency for the \texttt{Sh,Sv,Q6,Bh,Bv} Jacobian is twice the one for \texttt{S,Q3,B}.

\bibliographystyle{ieeetr}
\bibliography{literature}

\end{document}